\documentclass[11pt,oneside,english,american,reqno]{amsart}
\usepackage[T1]{fontenc}
\usepackage[latin9]{inputenc}
\usepackage{verbatim}
\usepackage{amstext}
\usepackage{amsthm}
\usepackage{amssymb}
\usepackage{undertilde}
\usepackage{xargs}[2008/03/08]

\makeatletter
\numberwithin{equation}{section}
\theoremstyle{plain}
\newtheorem{thm}{\protect\theoremname}[section]
  \theoremstyle{remark}
  \newtheorem{rem}[thm]{\protect\remarkname}

\usepackage{pdfsync} 
\usepackage{amscd}
\newcommand{\ie}{\textit{i.e.}}

\usepackage{natbib}
\usepackage[lining,tabular]{fbb} 
\usepackage[libertine,bigdelims]{newtxmath}
\usepackage[cal=boondoxo,bb=boondox,frak=boondox]{mathalfa}
\usepackage{diagrams}
\usepackage[all]{xy}
\makeatother

\usepackage{babel}
  \addto\captionsamerican{\renewcommand{\remarkname}{Remark}}
  \addto\captionsamerican{\renewcommand{\theoremname}{Theorem}}
  \addto\captionsenglish{\renewcommand{\remarkname}{Remark}}
  \addto\captionsenglish{\renewcommand{\theoremname}{Theorem}}
  \providecommand{\remarkname}{Remark}
\providecommand{\theoremname}{Theorem}

\begin{document}

\global\long\def\ga{\alpha}
\global\long\def\gb{\beta}
\global\long\def\ggm{\gamma}
\global\long\def\go{\omega}
\global\long\def\ge{\epsilon}
\global\long\def\gs{\sigma}
\global\long\def\gd{\delta}
\global\long\def\gD{\Delta}
\global\long\def\vph{\varphi}
\global\long\def\gf{\varphi}
\global\long\def\gk{\kappa}
\global\long\def\gG{\Gamma}

\global\long\def\eps{\varepsilon}
\global\long\def\epss#1#2{\varepsilon_{#2}^{#1}}
\global\long\def\ep#1{\eps_{#1}}

\global\long\def\wh#1{\widehat{#1}}

\global\long\def\spec#1{\textsf{#1}}

\global\long\def\ui{\wh{\boldsymbol{\imath}}}
\global\long\def\uj{\wh{\boldsymbol{\jmath}}}
\global\long\def\uk{\widehat{\boldsymbol{k}}}

\global\long\def\uI{\widehat{\mathbf{I}}}
\global\long\def\uJ{\widehat{\mathbf{J}}}
\global\long\def\uK{\widehat{\mathbf{K}}}

\global\long\def\bs#1{\boldsymbol{#1}}
\global\long\def\vect#1{\mathbf{#1}}
\global\long\def\bi#1{\textbf{\emph{#1}}}

\global\long\def\uv#1{\widehat{\boldsymbol{#1}}}
\global\long\def\cross{\times}

\global\long\def\ddt{\frac{\dee}{\dee t}}
\global\long\def\dbyd#1{\frac{\dee}{\dee#1}}
\global\long\def\dby#1#2{\frac{\partial#1}{\partial#2}}

\global\long\def\vct#1{\mathbf{#1}}

\global\long\def\partialby#1#2{\frac{\partial#1}{\partial x^{#2}}}
\newcommandx\parder[2][usedefault, addprefix=\global, 1=]{\frac{\partial#2}{\partial#1}}

\global\long\def\oneto{1,\dots,}
\global\long\def\mi#1{\boldsymbol{#1}}
\global\long\def\mii{\mi I}

\global\long\def\fall{,\quad\text{for all}\quad}

\global\long\def\reals{\mathbb{R}}

\global\long\def\rthree{\reals^{3}}
\global\long\def\rsix{\reals^{6}}
\global\long\def\rn{\reals^{n}}
\global\long\def\rt#1{\reals^{#1}}

\global\long\def\les{\leqslant}
\global\long\def\ges{\geqslant}

\global\long\def\dee{\textrm{d}}
\global\long\def\di{d}

\global\long\def\from{\colon}
\global\long\def\tto{\longrightarrow}
\global\long\def\lmt{\longmapsto}

\global\long\def\abs#1{\left|#1\right|}

\global\long\def\isom{\cong}

\global\long\def\comp{\circ}

\global\long\def\cl#1{\overline{#1}}

\global\long\def\fun{\varphi}

\global\long\def\interior{\textrm{Int}\,}

\global\long\def\sign{\textrm{sign}\,}
\global\long\def\sgn#1{(-1)^{#1}}
\global\long\def\sgnp#1{(-1)^{\abs{#1}}}

\global\long\def\dimension{\textrm{dim}\,}

\global\long\def\esssup{\textrm{ess}\,\sup}

\global\long\def\ess{\textrm{{ess}}}

\global\long\def\kernel{\mathop{\textrm{Kernel}}}

\global\long\def\support{\textrm{supp}\,}

\global\long\def\image{\textrm{Image}\,}

\global\long\def\diver{\mathop{\textrm{div}}}

\global\long\def\sdiv{\mathrm{div}}

\global\long\def\sp{\mathop{\textrm{span}}}

\global\long\def\resto#1{|_{#1}}
\global\long\def\incl{\iota}
\global\long\def\iden{\imath}
\global\long\def\idnt{\textrm{Id}}
\global\long\def\rest{\rho}
\global\long\def\extnd{e_{0}}

\global\long\def\proj{\textrm{pr}}

\global\long\def\ino#1{\int_{#1}}

\global\long\def\half{\frac{1}{2}}
\global\long\def\shalf{{\scriptstyle \half}}
\global\long\def\third{\frac{1}{3}}

\global\long\def\empt{\varnothing}

\global\long\def\paren#1{\left(#1\right)}
\global\long\def\bigp#1{\bigl(#1\bigr)}
\global\long\def\biggp#1{\biggl(#1\biggr)}
\global\long\def\Bigp#1{\Bigl(#1\Bigr)}

\global\long\def\braces#1{\left\{  #1\right\}  }
\global\long\def\sqbr#1{\left[#1\right]}
\global\long\def\anglep#1{\left\langle #1\right\rangle }

\global\long\def\lsum{{\textstyle \sum}}

\global\long\def\bigabs#1{\bigl|#1\bigr|}

\global\long\def\lisub#1#2#3{#1_{1}#2\dots#2#1_{#3}}

\global\long\def\lisup#1#2#3{#1^{1}#2\dots#2#1^{#3}}

\global\long\def\lisubb#1#2#3#4{#1_{#2}#3\dots#3#1_{#4}}

\global\long\def\lisubbc#1#2#3#4{#1_{#2}#3\cdots#3#1_{#4}}

\global\long\def\lisubbwout#1#2#3#4#5{#1_{#2}#3\dots#3\widehat{#1}_{#5}#3\dots#3#1_{#4}}

\global\long\def\lisubc#1#2#3{#1_{1}#2\cdots#2#1_{#3}}

\global\long\def\lisupc#1#2#3{#1^{1}#2\cdots#2#1^{#3}}

\global\long\def\lisupp#1#2#3#4{#1^{#2}#3\dots#3#1^{#4}}

\global\long\def\lisuppc#1#2#3#4{#1^{#2}#3\cdots#3#1^{#4}}

\global\long\def\lisuppwout#1#2#3#4#5#6{#1^{#2}#3#4#3\wh{#1^{#6}}#3#4#3#1^{#5}}

\global\long\def\lisubbwout#1#2#3#4#5#6{#1_{#2}#3#4#3\wh{#1}_{#6}#3#4#3#1_{#5}}

\global\long\def\lisubwout#1#2#3#4{#1_{1}#2\dots#2\widehat{#1}_{#4}#2\dots#2#1_{#3}}

\global\long\def\lisupwout#1#2#3#4{#1^{1}#2\dots#2\widehat{#1^{#4}}#2\dots#2#1^{#3}}

\global\long\def\lisubwoutc#1#2#3#4{#1_{1}#2\cdots#2\widehat{#1}_{#4}#2\cdots#2#1_{#3}}

\global\long\def\twp#1#2#3{\dee#1^{#2}\wedge\dee#1^{#3}}

\global\long\def\thp#1#2#3#4{\dee#1^{#2}\wedge\dee#1^{#3}\wedge\dee#1^{#4}}

\global\long\def\fop#1#2#3#4#5{\dee#1^{#2}\wedge\dee#1^{#3}\wedge\dee#1^{#4}\wedge\dee#1^{#5}}

\global\long\def\idots#1{#1\dots#1}
\global\long\def\icdots#1{#1\cdots#1}

\global\long\def\norm#1{\|#1\|}

\global\long\def\nonh{\heartsuit}

\global\long\def\nhn#1{\norm{#1}^{\nonh}}

\global\long\def\trps{^{{\scriptscriptstyle \textsf{T}}}}

\global\long\def\testfuns{\mathcal{D}}

\global\long\def\ntil#1{\tilde{#1}{}}

\global\long\def\alt{\mathfrak{A}}

\global\long\def\pou{\eta}

\global\long\def\ext{{\textstyle \bigwedge}}
\global\long\def\forms{\Omega}

\global\long\def\dotwedge{\dot{\mbox{\ensuremath{\wedge}}}}

\global\long\def\vel{\theta}

\global\long\def\flow{\phi}

\global\long\def\contr{\raisebox{0.4pt}{\mbox{\ensuremath{\lrcorner}}}\,}

\global\long\def\lie{\mathcal{L}}

\global\long\def\L#1{L\bigl(#1\bigr)}

\global\long\def\vvforms{\ext^{\dims}\bigp{T\spc,\vbts^{*}}}

\global\long\def\spc{\mathcal{S}}
\global\long\def\sptm{\mathcal{E}}
\global\long\def\evnt{e}
\global\long\def\frame{\Phi}

\global\long\def\timeman{\mathcal{T}}
\global\long\def\zman{t}
\global\long\def\dims{n}
\global\long\def\m{\dims-1}
\global\long\def\dimw{m}

\global\long\def\wc{z}

\global\long\def\fourv#1{\mbox{\ensuremath{\mathfrak{#1}}}}

\global\long\def\pbform#1{\utilde{#1}}
\global\long\def\util#1{\raisebox{-5pt}{\ensuremath{{\scriptscriptstyle \sim}}}\!\!\!#1}

\global\long\def\utilJ{\util J}

\global\long\def\utilRho{\util{\rho}}

\global\long\def\body{B}
\global\long\def\man{\mathcal{M}}
\global\long\def\var{\mathcal{V}}
\global\long\def\edg{\mathcal{E}}

\global\long\def\bdry{\partial}

\global\long\def\gO{\varOmega}

\global\long\def\reg{\mathcal{R}}
\global\long\def\bdrr{\bdry\reg}

\global\long\def\bdom{\bdry\gO}

\global\long\def\bndo{\partial\gO}

\global\long\def\pis{x}
\global\long\def\xo{\pis_{0}}

\global\long\def\pib{X}

\global\long\def\pbndo{\Gamma}
\global\long\def\bndoo{\pbndo_{0}}
 \global\long\def\bndot{\pbndo_{t}}

\global\long\def\cloo{\cl{\gO}}

\global\long\def\nor{\mathbf{n}}

\global\long\def\dA{\,\dee A}

\global\long\def\dV{\,\dee V}

\global\long\def\eps{\varepsilon}

\global\long\def\vs{\mathbf{W}}
\global\long\def\avs{\mathbf{V}}
\global\long\def\affsp{\mathbf{A}}
\global\long\def\pt{p}

\global\long\def\vbase{e}
\global\long\def\sbase{\mathbf{e}}
\global\long\def\msbase{\mathfrak{e}}
\global\long\def\vect{v}

\global\long\def\vf{w}

\global\long\def\avf{u}

\global\long\def\stn{\varepsilon}

\global\long\def\rig{r}

\global\long\def\rigs{\mathcal{R}}

\global\long\def\qrigs{\!/\!\rigs}

\global\long\def\qd{\!/\,\!\kernel\diffop}

\global\long\def\dis{\chi}
\global\long\def\conf{\kappa}

\global\long\def\fc{F}

\global\long\def\st{\sigma}

\global\long\def\bfc{\mathbf{b}}

\global\long\def\sfc{\mathbf{t}}

\global\long\def\stm{S}

\global\long\def\nhs{Y}

\global\long\def\soc{Z}

\global\long\def\ssp{\tau}
\global\long\def\sst{\tau}

\global\long\def\tran{\mathrm{tr}}

\global\long\def\slf{R}

\global\long\def\sts{\varSigma}

\global\long\def\ebdfc{T}
\global\long\def\optimum{\st^{\textrm{opt}}}
\global\long\def\scf{K}

\global\long\def\pform{\varsigma}
\global\long\def\vform{\beta}
\global\long\def\sform{\tau}
\global\long\def\n{\m}
\global\long\def\cmap{\mathfrak{t}}
\global\long\def\vcmap{\varSigma}

\global\long\def\mvec{\mathfrak{v}}
\global\long\def\mveco#1{\mathfrak{#1}}
\global\long\def\smbase{\mathfrak{e}}
\global\long\def\spx{\simp}

\global\long\def\hp{H}
\global\long\def\ohp{h}

\global\long\def\hps{G_{\dims-1}(T\spc)}
\global\long\def\ohps{G_{\dims-1}^{\perp}(T\spc)}
\global\long\def\hpsx{G_{\dims-1}(\tspc)}
\global\long\def\ohpsx{G_{\dims-1}^{\perp}(\tspc)}

\global\long\def\fbun{F}

\global\long\def\flowm{\Phi}

\global\long\def\tgb{T\spc}
\global\long\def\ctgb{T^{*}\spc}
\global\long\def\tspc{T_{\pis}\spc}
\global\long\def\dspc{T_{\pis}^{*}\spc}

\global\long\def\fflow{\fourv J}
\global\long\def\fvform{\mathfrak{b}}
\global\long\def\fsform{\mathfrak{t}}
\global\long\def\fpform{\mathfrak{s}}

\global\long\def\maxw{\mathfrak{g}}
\global\long\def\frdy{\mathfrak{f}}
\global\long\def\ptnl{A}

\global\long\def\eucl{E}

\global\long\def\mind{\alpha}
\global\long\def\vb{\xi}

\global\long\def\man{\mathcal{M}}
\global\long\def\odman{\mathcal{N}}
\global\long\def\subman{\mathcal{A}}

\global\long\def\vbt{\mathcal{E}}
\global\long\def\fib{\mathbf{V}}
\global\long\def\vbts{W}
\global\long\def\avb{U}

\global\long\def\chart{\varphi}
\global\long\def\vbchart{\Phi}

\global\long\def\jetb#1{J^{#1}}
\global\long\def\jet#1{j^{1}(#1)}

\global\long\def\Jet#1{J^{1}(#1)}

\global\long\def\jetm#1{j_{#1}}

\global\long\def\conn{\mathcal{C}}
\global\long\def\jete{\mathcal{E}}
\global\long\def\norp{\mathcal{N}}
\global\long\def\vbf{\mathcal{F}}

\global\long\def\sobp#1#2{W_{#2}^{#1}}

\global\long\def\inner#1#2{\left\langle #1,#2\right\rangle }

\global\long\def\fields{\sobp pk(\vb)}

\global\long\def\bodyfields{\sobp p{k_{\partial}}(\vb)}

\global\long\def\forces{\sobp pk(\vb)^{*}}

\global\long\def\bfields{\sobp p{k_{\partial}}(\vb\resto{\bndo})}

\global\long\def\loadp{(\sfc,\bfc)}

\global\long\def\strains{\lp p(\jetb k(\vb))}

\global\long\def\stresses{\lp{p'}(\jetb k(\vb)^{*})}

\global\long\def\diffop{D}

\global\long\def\strainm{E}

\global\long\def\incomps{\vbts_{\yieldf}}

\global\long\def\devs{L^{p'}(\eta_{1}^{*})}

\global\long\def\incompsns{L^{p}(\eta_{1})}

\global\long\def\testf{\mathcal{D}}
\global\long\def\dists{\mathcal{D}'}

\global\long\def\codiv{\boldsymbol{\partial}}

\global\long\def\currof#1{\tilde{#1}}

\global\long\def\chn{c}
\global\long\def\chnsp{\mathbf{F}}

\global\long\def\current{T}
\global\long\def\curr#1{T_{\langle#1\rangle}}

\global\long\def\prop{P}

\global\long\def\aprop{Q}

\global\long\def\flux{T}
\global\long\def\aflux{S}

\global\long\def\fform{\tau}

\global\long\def\dimn{n}

\global\long\def\sdim{{\dimn-1}}

\global\long\def\contrf{{\scriptstyle \smallfrown}}

\global\long\def\prodf{{\scriptstyle \smallsmile}}

\global\long\def\ptnl{\varphi}

\global\long\def\form{\omega}

\global\long\def\dens{\rho}

\global\long\def\simp{s}
\global\long\def\ssimp{\Delta}
\global\long\def\cpx{K}

\global\long\def\cell{C}

\global\long\def\chain{B}

\global\long\def\ach{A}

\global\long\def\coch{X}

\global\long\def\scale{s}

\global\long\def\fnorm#1{\norm{#1}^{\flat}}

\global\long\def\chains{\mathcal{A}}

\global\long\def\ivs{\boldsymbol{U}}

\global\long\def\mvs{\boldsymbol{V}}

\global\long\def\cvs{\boldsymbol{W}}

\global\long\def\cee#1{C^{#1}}

\global\long\def\lone{L^{1}}

\global\long\def\linf{L^{\infty}}

\global\long\def\lp#1{L^{#1}}

\global\long\def\ofbdo{(\bndo)}

\global\long\def\ofclo{(\cloo)}

\global\long\def\vono{(\gO,\rthree)}

\global\long\def\vonbdo{(\bndo,\rthree)}
\global\long\def\vonbdoo{(\bndoo,\rthree)}
\global\long\def\vonbdot{(\bndot,\rthree)}

\global\long\def\vonclo{(\cl{\gO},\rthree)}

\global\long\def\strono{(\gO,\reals^{6})}

\global\long\def\sob{W_{1}^{1}}

\global\long\def\sobb{\sob(\gO,\rthree)}

\global\long\def\lob{\lone(\gO,\rthree)}

\global\long\def\lib{\linf(\gO,\reals^{12})}

\global\long\def\ofO{(\gO)}

\global\long\def\oneo{{1,\gO}}
\global\long\def\onebdo{{1,\bndo}}
\global\long\def\info{{\infty,\gO}}

\global\long\def\infclo{{\infty,\cloo}}

\global\long\def\infbdo{{\infty,\bndo}}

\global\long\def\ld{LD}

\global\long\def\ldo{\ld\ofO}
\global\long\def\ldoo{\ldo_{0}}

\global\long\def\trace{\gamma}

\global\long\def\pr{\proj_{\rigs}}

\global\long\def\pq{\proj}

\global\long\def\qr{\,/\,\reals}

\global\long\def\aro{S_{1}}
\global\long\def\art{S_{2}}

\global\long\def\mo{m_{1}}
\global\long\def\mt{m_{2}}

\global\long\def\yieldc{B}

\global\long\def\yieldf{Y}

\global\long\def\trpr{\pi_{P}}

\global\long\def\devpr{\pi_{\devsp}}

\global\long\def\prsp{P}

\global\long\def\devsp{D}

\global\long\def\ynorm#1{\|#1\|_{\yieldf}}

\global\long\def\colls{\Psi}

\global\long\def\ssx{S}

\global\long\def\smap{s}

\global\long\def\smat{\chi}

\global\long\def\sx{e}

\global\long\def\snode{P}

\global\long\def\elem{e}

\global\long\def\nel{L}

\global\long\def\el{l}

\global\long\def\ipln{\phi}

\global\long\def\ndof{D}

\global\long\def\dof{d}

\global\long\def\nldof{N}

\global\long\def\ldof{n}

\global\long\def\lvf{\chi}

\global\long\def\lfc{\varphi}

\global\long\def\amat{A}

\global\long\def\snomat{E}

\global\long\def\femat{E}

\global\long\def\tmat{T}

\global\long\def\fvec{f}

\global\long\def\snsp{\mathcal{S}}

\global\long\def\slnsp{\Phi}

\global\long\def\ro{r_{1}}

\global\long\def\rtwo{r_{2}}

\global\long\def\rth{r_{3}}

\global\long\def\subbs{\mathcal{B}}

\global\long\def\elements{\mathcal{E}}

\global\long\def\element{E}

\global\long\def\nodes{\mathcal{N}}

\global\long\def\node{N}

\global\long\def\psubbs{\mathcal{P}}

\global\long\def\psubb{P}

\global\long\def\matr{M}

\global\long\def\nodemap{\nu}

\global\long\def\node{v}

\global\long\def\edge{e}

\global\long\def\accu{q}

\global\long\def\accusp{\mathcal{Q}}

\global\long\def\potl{\varphi}

\global\long\def\ptnl{\alpha}

\global\long\def\currsp{\mathcal{I}}

\global\long\def\volt{V}

\global\long\def\intv{\mathbf{t}}
\global\long\def\intc{t}
\global\long\def\intsp{\mathcal{T}}

\global\long\def\frcv{\mathbf{f}}
\global\long\def\frcc{f}
\global\long\def\frcsp{\mathcal{F}}

\global\long\def\velv{\mathbf{V}}
\global\long\def\velc{V}
\global\long\def\disv{\mathbf{E}}
\global\long\def\disc{E}

\global\long\def\posn{\mathbf{x}}
\global\long\def\area{\mathbf{A}}
\global\long\def\relp{\mathbf{L}}

\global\long\def\chn{c}

\title[Geometric Analysis of Hyper-Stresses]{}

\title{Geometric Analysis of Hyper-Stresses}

\author{Reuven Segev}

\curraddr{Reuven Segev\\
Department of Mechanical Engineering\\
Ben-Gurion University of the Negev\\
Beer-Sheva, Israel\\
rsegev@bgu.ac.il}


\thanks{\today}

\begin{abstract}
A geometric analysis of high order stresses in continuum mechanics
is presented. Virtual velocity fields take their values in a vector
bundle $\vbts$ over the $n$-dimensional space manifold. A stress
field of order $k$ is represented mathematically by an $n$-form
valued in the dual of the vector bundle of $k$-jets of $\vbts$.
While only limited analysis can be performed on high order stresses
as such, they may be represented by non-holonomic hyper-stresses,
$n$-forms valued in the duals of iterated jet bundles. For non-holonomic
hyper-stresses, the analysis that applies to first order stresses
may be iterated. In order to determine a unique value for the tangent
surface stress field on the boundary of a body and the corresponding
edge interactions, additional geometric structure should be specified,
that of a vector field transversal to the boundary. 
\end{abstract}

\maketitle

\section{Introduction}

The theory of hyper-stresses in continuum mechanics, \emph{e.g.} \cite{Toupin62,Toupin64,Mindlin64,Mindlin65},
accounts for phenomena not accounted for by the standard theory of
stresses, such as edge interactions and surface tension. Although
five decades have passed since this pioneering body of work has been
published, various aspects of higher-order continuum mechanics are
still under current research, \emph{e.g.} \cite{dellIsola2012,dellIsoll2015,Fosdick2016,Mariano2007,Neff2016,PPG2015}.

This work is concerned with geometric analysis of smooth stresses
of order $k$ in continuum mechanics. In \cite{Segev1986}, for the
setting where both the body $\body$ and space $\spc$ objects of
continuum mechanics are modeled as general differentiable manifolds,
a hyper-stress theory was proposed in which the fundamental object
is the configuration space $Q$ containing all $C^{k}$-embeddings
of the body into space. Using results on manifolds of mappings (\emph{e.g.}
\cite{Palais68,Michor1980,Hirsch}), it follows that the configuration
space may be given the structure of a Banach manifold. The tangent
space $T_{\conf}Q$, at a generic configuration of the body $\conf:\body\to\spc$,
is interpreted physically as the space of virtual velocities. It may
be identified with the space of $C^{k}$-sections, vector fields,
of some vector bundle $\vbts$, where the space of sections is equipped
with the $C^{k}$-topology. A generalized force $\fc$ of order $k$
at the configuration $\conf$ is defined to be a continuous linear
functional on the tangent space $T_{\conf}Q$ and the value of the
action of a force on a generalized velocity is interpreted as the
corresponding virtual power.

It is shown there that forces may be represented by measures valued
in the dual of the $k$-jet bundle, $J^{k}\vbts$, of $\vbts$. Locally,
these measures are represented by a collection of tensors valued measures
of orders $1$ to $k$. These representing measures are referred to
as variational (hyper-) stresses. The relation between a force system
containing the forces of order $k$ on all subbodies of $\body$ and
a $k$-hyper-stress field, the analog of Cauchy's postulates, is studied
in \cite{Segev1986,SegevDeBotton1991} for the general case of stress
fields that are as irregular as measures.

In the smooth case, the measures of the variational stress are represented
by smooth sections $\stm$ of the fiber bundle $\L{J^{k}\vbts,\ext^{n}T^{*}\body}=(J^{k}\vbts)^{*}\otimes\ext^{n}T^{*}\body$
so that the value of the stress field at a point $x\in\body$ is a
linear mapping $(J^{k}\vbts)_{x}\to\ext^{n}T^{*}\body$. Thus, the
power expended by the force $\fc$ for the generalized velocity $\vf$
is given by
\begin{equation}
\fc(\vf)=\int_{\body}\stm(j^{k}\vf),\label{eq:IntStressRep}
\end{equation}
where $\stm(j^{k}\vf)$ is the $n$-form whose value at $x\in\body$
is $\stm(x)(j^{k}\vf(x))$, so that the integration above is well
defined.

For the standard continuum mechanics case, $k=1$, a procedure given
in \cite{Segev2002,Segev2013} and outlined in Section \ref{sec:Simple-Stresses},
makes is possible to write (\ref{eq:IntStressRep}) in the form
\begin{equation}
\fc(\vf)=\int_{\body}\bfc(\vf)+\int_{\bdry\body}\sfc(\vf),
\end{equation}
where $\bfc$, the body force, is a section of $\L{\vbts,\ext^{n}T^{*}\body}$,
satisfies
\begin{equation}
\diver\stm+\bfc=0,\quad\text{in }\body,
\end{equation}
and $\sfc$, the surface force, satisfies a generalization of Cauchy's
formula 
\begin{equation}
\rho\comp\st=\sfc,\quad\text{on }\bdry\body.
\end{equation}
Here, $\st$, the traction stress, is a section of $\L{\vbts,\ext^{n-1}T^{*}\body}$
that generalizes the Cauchy stress, and $\rho$ is the restriction
of forms defined on $T\body$ to $T\bdry\body$. The traction stress
is determined by the variational stress. It is emphasized that for
the setting of general manifolds, two distinct objects represent the
two functions of the classical stress object, namely, acting on derivative
of velocities to produce power, and determining the surface force
for various subbodies. The divergence operator for manifolds, as mentioned
above and defined below, generalizes the standard divergence operator
of second order tensors.

In this paper we study the geometric structure required to provide
the analogous construction for smooth hyper-stresses of order $k$.
In particular, we consider the geometric structure needed to determine
the edge interactions induced by hyper-stresses using integral transformations
in analogy with the analysis in \cite{dellIsola2012,dellIsoll2015}.
It is shown below that the setting of iterated jet bundles, $J^{1}(J^{1}\vbts)$
for instance, is preferable to that of higher jet bundles, for instance,
$J^{2}\vbts$, respectively. Using iterated jet bundles makes it possible
to apply the procedure for standard continuum mechanics, inductively.

Hyper-stresses in bodies induce tangent surface stresses on the corresponding
boundaries. However, it is shown that on general differentiable manifolds,
the induced surface stress, and hence the edge interactions, are not
unique. For the unique determination of the tangent surface stress,
one needs at least some specified vector field which is transversal
to the boundary or an equivalent structure. The situation is similar
to that described in \cite{Epstein73,Epstein98}, where shell theory
is considered. Evidently, for the case of a Riemannian manifold, the
unit normal vector field provides such a transversal field naturally.

Section \ref{sec:Notation-and-Prliminaries} introduces the relevant
terminology and notation used for jet bundles associated with vector
bundles. Section \ref{sec:Simple-Stresses} reviews the relevant constructions
of \cite{Segev2002} regarding smooth stress distributions on manifolds
as outlined above. Section \ref{sec:High-Order-Stresses} is concerned
with hyper-stresses of order $k$, their representations and their
invariant components. Some of the difficulties related to the analysis
of hyper-stresses are indicated. Section \ref{sec:Iterated-Jet-Bundle}
considers iterated jet bundles (see \cite{Saunders}). Iterated jet
bundles are of interest as their sections may have additional forms
of incompatibility in comparison with sections of jet bundles. Forms
valued in the duals of iterated jet bundles are referred to here as
non-holonomic hyper-stresses. These are considered in Section \ref{sec:Non-Holonomic-Stresses}.
Due to the inductive nature of iterated jet bundles, it is sufficient
to study the properties of the iterated jet bundle $J^{1}(J^{1}\vbts)$.
The vector bundle $\vbts$ itself may be a jet bundle, or an iterated
jet bundle, of some other vector bundle. It is noted that every hyper-stress
may be represented by non-holonomic hyper-stresses. The properties
of non-holonomic hyper-stresses, in particular, the corresponding
integral transformations associated with their action, are analyzed
in this section for the case of general manifolds. Section \ref{sec:Additional-Geometric-Structure}
shows how the introduction of a particular vector field which is transversal
to the boundary of a body induces a unique stress of a lower order
on the boundary. Finally, in Section \ref{sec:Edge-Interactions},
the edge interactions induced by the non-holonomic hyper-stress are
computed.

\section{Notation and Preliminaries\label{sec:Notation-and-Prliminaries}}

All manifolds considered here are viewed as chains or manifolds with
corners so that we may use the Stokes theorem for integration of forms.

\subsection{Jets in general}

We will use the same scheme of notation as in \cite{Segev2013} and
we will ofter use the same notation for a mapping and variables in
the co-domains thereof. Let $\pi:\vbts\to\spc$ be a vector bundle,
a section $\vf:\spc\to\vbts$ of $\pi$ is represented locally in
the form 
\begin{equation}
(\lisup x,{\dims})\longmapsto(\lisup x,n,\vf^{1}(x^{i}),\dots,\vf^{d}(x^{i})),\label{eq:ReprSectVB-1}
\end{equation}
where $(\lisup x,n)$ is a local coordinate system, and a local basis
$\{g_{1},\dots,g_{d}\}$ was used for the fibers of $\vbts$. Let
$\mii=(i_{1},\dots,i_{n})$, for non-negative integers $i_{j}$, be
a multi-index and let $\abs{\mii}=\sum_{j=1}^{n}i_{j}$. We use the
notation
\begin{equation}
\parder[x^{\mii}]{^{\abs{\mii}}}=\parder[x^{i_{1}}\cdots\bdry x^{i_{n}}]{^{\abs{\mii}}}.
\end{equation}

Two sections $\vf$ and $\vf'$ have the same \emph{$k$}-jet at \emph{$x_{0}\in\spc$}
if 
\begin{equation}
\parder[x^{\mii}]{^{\abs{\mii}}\vf^{\ga}}(x_{0}^{i})=\parder[x^{\mii}]{^{\abs{\mii}}\vf'^{\ga}}(x_{0}^{i})
\end{equation}
for all $\mii$ such that $\abs{\mii}\les k$ and all $\ga=1,\dots,d$.
Clearly, if this condition holds in one vector bundle chart in a neighborhood
of $x_{0}$, it will hold in any other chart and it induces an equivalence
relation on the vector space $C^{k}(\vbts)=C^{k}(\pi)$ of $C^{k}$-sections
of the vector bundle. An equivalence class for this relation is a\emph{
$k$}-jet at $x_{0}$. Given a section $\vf$, the jet it induces
at $x_{0}$\textemdash the jet of\emph{ $w$ }at $x_{0}$\textemdash will
be denoted as $j^{k}(w)(x_{0})$. Given a chart in a neighborhood
of $x_{0}$, $j^{k}(\vf)(x_{0})$ is represented by 
\begin{equation}
\left\{ \vf_{,\mii}^{\ga}(x_{0}):=\parder[x^{\mii}]{^{\abs{\mii}}\vf^{\ga}}(x_{0}^{i})\mid\abs{\mii}\les k,\,\ga=1,\dots,d\right\} .\label{eq:ReprOfaJet}
\end{equation}

The collection of all $k$-jets at $x_{0}\in\spc$ is the \emph{$k$}-jet
space of the vector bundle at $x_{0}$ and is denoted as $J_{x_{0}}^{k}\vbts$.
The $k$-jet bundle $J^{k}\vbts$ is the collection of all $k$-jets
at the various points in $\spc$ so that 
\begin{equation}
J^{k}\vbts=\bigcup_{x\in\spc}J_{x}^{k}\vbts.
\end{equation}
By convention, $J^{0}\vbts=\vbts$. A natural vector bundle structure
\begin{equation}
\pi^{k}:J^{k}\vbts\tto\spc,
\end{equation}
is available on the jet bundle by which $\pi^{k}(A)=x$ if $A\in J_{x}^{k}\vbts$.
The linear structure on the fibers is given by $a_{1}A_{1}+a_{2}A_{2}=j^{k}(a_{1}\vf_{1}+a_{2}\vf_{2})(x)$,
for $A_{1}$, $A_{2}$ in $J_{x}^{k}\vbts$, $a_{1},a_{2}\in\reals$,
and representing sections $\vf_{1}$ and $\vf_{2}$. Evidently, the
result is independent of the choice of representative sections. The
fiber $J_{x}^{k}\vbts$ of this vector bundle over $x\in\spc$ is
isomorphic with 
\begin{equation}
\vbts_{x}\oplus\L{T_{x}\spc,\vbts_{x}}\oplus\cdots\oplus L_{S}^{p}(T_{x}\spc,\vbts_{x})\oplus\cdots\oplus L_{S}^{k}(T_{x}\spc,\vbts_{x}),\label{eq:FiberOfJetBundle}
\end{equation}
where $L_{S}^{p}(T_{x}\spc,\vbts_{x})$ denotes the vector space of
$p$-multilinear symmetric mappings from $T_{x}\spc$ to $\vbts_{x}$.
Thus, an element in $J_{x}^{k}\vbts$ is represented locally in the
form
\begin{equation}
(A^{0\ga_{0}},A_{\mii_{1}}^{1\ga_{1}},\dots,A_{\mii_{p}}^{p\ga_{p}},\dots,A_{\mii_{k}}^{k\ga_{k}})=(A_{\mii}^{p\ga}),\label{eq:repJet}
\end{equation}
where $p=0,\dots,k$, $\abs{\mii_{p}}=p$, $\ga=(\ga_{0},\dots,\ga_{k})$,
$\ga_{p}=1,\dots,d$, and evidently, $A^{0\ga_{0}}$ represents an
element of $\vbts_{x}$. Each section $\vf$ of $\vbts$ induces a
section $j^{k}\vf$ of the $k$-th jet bundle and if fact we have
a continuous linear injection
\begin{equation}
j^{k}:C^{k}(\vbts)\tto C^{0}(J^{k}\vbts),
\end{equation}
where $C^{p}(\avb)$ represents the vector space of sections of the
vector bundle $\avb$ of class $p$. For additional information on
jet bundles, some of which will be used in the following sections,
see \cite{Saunders}.

A jet bundle has also the natural projections
\begin{equation}
\pi_{p}^{k}:J^{k}W\tto J^{p}\vbts,\quad0\les p\les k,
\end{equation}
characterized by $\pi_{p}^{k}(A)=j^{p}(\vf)(x)$ where $x=\pi^{k}(A)$
and $\vf$ is any section of $\vbts$ that represents $A$. The mapping
$\pi_{p}^{k}$ is a vector bundle morphism over $\spc$.

Let $\phi:\vbts\to\avb$ be a fiber bundle morphism over the base
manifold $\spc$. For an element $A\in J^{k}\vbts$, represented by
$j^{k}\vf(x)$, where $\vf$ is a section of $\vbts$, set $j^{k}\phi(A)=j^{k}(\phi\comp\vf)(x)\in J^{k}\avb$.
In this way, one defines the $k$-lift of $\phi$, the vector bundle
morphism
\begin{equation}
j^{k}\phi:J^{k}\vbts\tto J^{k}\avb.\label{eq:Lift_of_VBM}
\end{equation}
Evidently,
\begin{equation}
\pi_{p}^{k}\comp j^{k}\phi=j^{p}\phi\comp\pi_{r}^{k}.
\end{equation}

\subsection{Vertical sub-bundles}

For $0\les r<k$, we say that $A\in J^{k}\avb\resto x$ is $r$-vertical
if for one (and hence any) section $\avf$ representing $A$, $j^{r}\avf(x)=0$,
or equivalently, if $\pi_{r}^{k}(A)=0$. If $A$ is $r$-vertical,
then, its local representatives satisfy $A_{\mii}^{p\ga}=0$ for all
$p=\abs{\mii}\les r$. Thus, locally
\begin{equation}
V^{r}J^{k}\avb\resto x\isom L_{S}^{r+1}(T_{x}\spc,\avb_{x})\oplus\cdots\oplus L_{S}^{k}(T_{x}\spc,\avb_{x}).\label{eq:rep_verticals}
\end{equation}

The collection of $r$-vertical elements is a vector sub-bundle of
the jet bundle and we denote it by $V^{r}J^{k}\avb$, \ie, $V^{r}J^{k}\avb=\kernel\pi_{r}^{k}$.
One has the natural vector bundle inclusion 
\begin{equation}
\incl_{V}^{r}:V^{r}J^{k}\avb\to J^{k}\avb.\label{eq:Incl_Verts}
\end{equation}
For the particular case of the first jet bundle $J^{1}\avb$, the
only vertical sub-bundle is $V^{0}J^{1}\avb$, and  we will often
omit the zero superscript and write just $VJ^{1}\avb$.

For the case, $r=k-1$, one has a natural isomorphism
\begin{equation}
V^{k-1}J^{k}\avb\resto x\isom L_{S}^{k}(T_{x}\spc,\avb_{x}),\qquad V^{k-1}J^{k}\avb\isom L_{S}^{k}(T\spc,\avb).
\end{equation}
The vertical subbundle $V^{k-1}J^{k}\avb$ will be referred to as
the completely vertical sub-bundle of the $k$-jet bundle. In the
particular case $k=1$, $r=0$, it follows that $VJ^{1}\avb$ is naturally
isomorphic with $L(T\spc,\avb)$.
\begin{rem}
In view of (\ref{eq:rep_verticals}), one may be tempted to view elements
of $V^{r}J^{k}U$ as elements of $J^{k-r-1}(L_{S}^{r+1}(T\spc,\avb))$.
However, it may be easily verified that there is no such invariant
correspondence.
\end{rem}

\subsection{Some details on 1-jets\label{subsec:details-on-jets}}

Let $\vf:\spc\to\vbts$ be a section of $\pi:\vbts\to\spc$, then,
$j^{1}\vf(x)$, $x\in\spc$, is represented locally in the form $(x^{i},\vf^{\ga}(x),\vf_{,j}^{\gb}(x))$,
and $\pi_{0}^{1}(j^{1}\vf(x))$ is obviously $\vf(x)$ which is represented
locally in the form $(x^{i},\vf^{\ga})$. The tangent at $x$ to the
section $\vf$, $T_{x}\vf:T_{x}\spc\to T_{\vf(x)}\vbts$, is represented
locally by 
\begin{equation}
(x^{i},v^{j})\lmt\paren{x^{i},\vf^{\ga}(x),v^{j},{\textstyle \sum_{k}}\vf_{,k}^{\gb}v^{k}}.\label{eq:Rep_TyW}
\end{equation}
Since $T\pi:T\vbts\to T\spc$ is represented locally by $(x^{i},\vf^{\ga},\dot{x}^{j},\dot{\vf}^{\gb})\mapsto(x^{i},\dot{x}^{j})$,
 any linear mapping $\tilde{A}:T_{x}\spc\to T_{\vf(x)}\vbts$ satisfying
the condition $T\pi\comp\tilde{A}=\idnt$, induces a unique element
$A\in J^{1}\vbts_{x}$ with $\pi_{0}^{1}(A)=\vf(x)$. 

Let $\var$ be a submanifold of $\spc$ and for the natural embedding
$\incl_{\var}:\var\to\spc$, let $T\incl_{\var}:T\var\to T\spc$ be
the tangent mapping. Thus, with $i,j=1,\dots,n$ and $a,b=1,\dots,\dimension\var$,
$\incl_{\var}$ is represented in the form $(y^{a})\mapsto(\incl_{\var}^{i}(y^{a}))$
and $T\incl_{\var}$ is represented by $(y^{a},\dot{y}^{b})\mapsto\paren{\incl_{\var}^{i}(y^{a}),{\textstyle \sum_{b}}\incl_{\var,b}^{j}\dot{y}^{b}}$.
One has the pullback $\incl_{\var}^{*}\pi:\incl_{\var}^{*}\vbts\to\var$,
of the vector bundle $\pi$ onto $\var$, the natural inclusion $\pi^{*}\incl_{\var}:\incl_{\var}^{*}\vbts\to\vbts$,
its tangent $T(\pi^{*}\incl_{\var}):T(\incl_{\var}^{*}\vbts)\to T\vbts$,
the mapping $\incl_{\var\pi}^{*}:C^{1}(\pi)\to C^{1}(\incl_{\var}^{*}(\pi))$\textemdash which
is simply the restriction of sections of $\pi$ to $\var$, and the
corresponding jet bundle $\pi^{1}(\incl_{\var}^{*}\pi):J^{1}(\incl_{\var}^{*}\vbts)\to\var$.
Thus, we will often use the notation $\vbts\resto{\var}$ for the
pullback. Locally, $\pi^{*}\incl_{\var}$ is represented in the form
$(y^{a},u^{\ga})\mapsto(\incl^{i}(y^{a}),u^{\ga})$, and $T(\pi^{*}\incl_{\var}$)
is represented in the form $(y^{a},u^{\ga},\dot{y}^{b},\dot{u}^{\gb})\mapsto\paren{\incl^{i}(y^{a}),u^{\ga},\sum_{b}\incl_{\var,b}^{j}\dot{y}^{b},\dot{u}^{\gb}}$.
Similarly, one may consider the pullback $\incl_{\var}^{*}(\pi^{1}):\incl_{\var}^{*}(J^{1}\vbts)\to\var$
with the natural inclusion $\pi^{1*}(\incl_{\var}):\incl_{\var}^{*}(J^{1}\vbts)\to J^{1}\vbts$.

There is a natural restriction mapping $\rho=j^{1}\comp\incl_{\var\pi}^{*}:\incl_{\var}^{*}(J^{1}\vbts)\to J^{1}(\incl_{\var}^{*}\vbts)$
whereby $j^{1}\vf(y)\mapsto j^{1}(\incl_{\var\pi}^{*}\vf)(y)$, $y\in\var$,
and it is noted that $j^{1}$ on the right is the jet extension on
the submanifold $\var$ which we may also write as $j_{\var}^{1}$.

\section{Simple Stresses\label{sec:Simple-Stresses}}

As a primitive mathematical object pertaining to stress theory for
continuum mechanics of order 1 we take the variational stress, a smooth
section $\stm$ of the vector bundle $L(J^{1}\vbts,\ext^{n}T^{*}\spc)$
for some vector bundle $\vbts\to\spc$, where $\ext^{n}T^{*}\spc$
is the vector bundle of $n$-alternating covariant tensors over $\spc$.
For motivation, see \cite{Segev1986,Segev2002,Segev2013}. In particular,
for an $n$-dimensional submanifold with boundary $\body\subset\spc$,
one is interested in the linear functional, the force,
\begin{equation}
\fc_{\body}:\vf\longmapsto\int_{\body}\stm(j^{1}\vf)
\end{equation}
which is interpreted as the virtual power performed by the variational
stress $\stm$ for the virtual generalized velocity field $\vf$ inside
the region $\body$. Here, the jet extension of $\vf$ generalizes
the traditional gradient to the setting of differentiable manifolds.

Locally, $\stm$ is represented in the form $(x^{i},\stm_{1\dots n\ga}^{0},S_{1\dots n\ga}^{1j})$,
or in detail, denoting the natural base vectors induced by a chart
as $\partial_{i}=\bdry/\bdry x^{i}$, the local representation is
\begin{equation}
\left(\sum_{\ga}\stm_{1\dots n\ga}^{0}+\sum_{i,\ga}\stm_{1\dots n\ga}^{1i}\otimes\partial_{i}\right)\otimes g^{\ga}\otimes(\lisupc{\dee x}{\wedge}n).
\end{equation}
Consequently, $\stm(j(\vf))$ is represented locally by 
\begin{equation}
\left(\sum_{\ga}\stm_{1\dots n\ga}^{0}\vf^{\ga}+\sum_{i,\ga}\stm_{1\dots n\ga}^{1i}\vf_{,i}^{\ga}\right)\lisupc{\dee x}{\wedge}n.\label{eq:ReprPowerIntern-1}
\end{equation}

For a vector bundle $V\to\spc$, let $\ext^{p}(T^{*}\spc,V)$ denote
the bundle of $V$-valued $p$-forms, \ie,  the vector bundle over
$\spc$ whose fiber at $x$ is the vector space of $p$-alternating
multilinear mappings from $T_{x}\spc$ to $V_{x}$. Consider the isomorphism
\begin{equation}
\tran:\ext^{p}(T^{*}\spc,V^{*})\tto L(V,\ext^{p}T^{*}\spc)
\end{equation}
defined as follows. For $T\in\ext^{p}(T^{*}\spc,V^{*})$, $T^{\tran}=\tran(T)$
is given by 
\begin{equation}
T^{\tran}(v)(\lisub u,p)=T(\lisub u,p)(v).
\end{equation}
Thus, for a variational stress $\stm$ one may consider $\stm\trps=\tran^{-1}(\stm)$\textemdash an
$n$-form on $\spc$ valued in the dual of the jet bundle.

\subsection{Traction stresses\label{subsec:Traction-stresses}}

Consider the inclusion $\incl_{V}:VJ^{1}\vbts\to J^{1}\vbts$. Then,
the dual vector bundle morphism $\incl_{V}^{*}:(J^{1}\vbts)^{*}\to(VJ^{1}\vbts)^{*}\isom\L{\vbts,T\spc}$
is a projection represented locally in the form $(x^{i},r_{p},R_{q}^{i})\mapsto(x^{i},R_{q}^{i})$\textemdash the
restriction of $R\in(J^{1}\vbts)^{*}$ to vertical elements of the
jet bundle. Thus, $\incl_{V}^{*}(R)(A)$, $A\in VJ^{1}\vbts$, is
represented by $\sum_{i,q}R_{q}^{i}A_{i}^{q}$. Similarly, for a section
$\stm$ of $\L{J^{1}\vbts,\ext^{n}T^{*}\spc}$, $\incl_{V}^{*}(\stm):=\incl_{V}^{*}\comp\stm$,
a section of $\L{VJ^{1}\vbts,\ext^{d}T^{*}\spc}$, is given by $\incl_{V}^{*}(\stm)(x)(A)=\stm(x)(\incl_{V}(A))\in\ext^{n}T_{x}^{*}\spc$.
The evaluation $\incl_{V}^{*}(\stm)(x)(A)$ is represented by $\sum_{j,\ga}\stm_{1\dots n\ga}^{1j}(x)A_{j}^{\ga}\lisupc{\dee x}{\wedge}n$
and so $\incl_{V}^{*}(\stm)$ is represented in the form w
\begin{equation}
\sum_{j,\ga}\stm_{1\dots n\ga}^{1j}\partial_{j}\otimes g^{\ga}\otimes(\lisupc{\dee x}{\wedge}n).\label{eq:RepresSymbol-1}
\end{equation}
The object $\incl_{V}^{*}(\stm)$ is the symbol of the linear differential
operator $\stm$ as defined in \cite{Palais68}.

Using the isomorphism $VJ^{1}\vbts\isom\L{T\spc,\vbts}$, we view
$\incl_{V}^{*}(\stm)$ as a section of 
\begin{equation}
\begin{split}\L{\L{T\spc,\vbts},\ext^{n}T^{*}\spc} & \isom\L{T\spc,\vbts}^{*}\otimes\ext^{n}T^{*}\spc,\\
 & \isom\L{\vbts,T\spc}\otimes\ext^{n}T^{*}\spc,\\
 & \isom\vbts^{*}\otimes T\spc\otimes\ext^{n}T^{*}\spc.
\end{split}
\end{equation}
It follows that a section of $\ext^{\dims}(T^{*}\spc,L(\vbts,T\spc))$
may be represented locally in the form $\sum_{a}\fun^{a}\otimes v_{a}\otimes\theta$
for an $n$-form $\theta$ and pairs $v_{a},$ $\fun^{a}$ of sections
of $T\spc$ and $\vbts^{*}$, respectively. We can use the contraction
of the second and first factors in the product to obtain $\sum_{a}\fun^{a}\otimes(v_{a}\contr\theta)$.
Thus, we have a natural mapping
\begin{equation}
\begin{split}\spec C:\L{\L{T\spc,\vbts},\ext^{n}T^{*}\spc} & \tto\vbts^{*}\otimes\ext^{n-1}T^{*}\spc\\
 & \,\,\,\,\,\isom\L{\vbts,\ext^{n-1}T^{*}\spc}.
\end{split}
\label{eq:contraction-map}
\end{equation}
The mapping $\spec C$ is represented locally by
\begin{multline}
\sum_{j,\ga}\stm_{1\dots n\ga}^{1j}\bdry_{j}\otimes g^{\ga}\otimes(\lisupc{\dee x}{\wedge}n)\longmapsto\sum_{j,\ga}\stm_{1\dots n\ga}^{1j}g^{\ga}\otimes\left(\partial_{j}\contr(\lisupc{\dee x}{\wedge}n)\right),\\
=\sum_{j,\ga}(-1)^{j-1}\stm_{1\dots n\ga}^{1j}g^{\ga}\otimes(\lisupwout{\dee x}{\wedge}nj),\label{eq:contraction-defined}
\end{multline}
where a superimposed ``hat'' indicates the omission of the specified
term.

The mapping 
\begin{equation}
p_{\st}:=\spec C\comp\incl_{V}^{*}:\L{\L{T\spc,\vbts},\ext^{n}T^{*}\spc}\tto\L{\vbts,\ext^{n-1}T^{*}\spc}\label{eq:ProjectionToTractionSt}
\end{equation}
associates a section $\st=p_{\st}\comp\stm$ of $\L{\vbts,\ext^{n-1}T^{*}\spc}$
with a variational stress $\stm$. We refer to a section of $\L{\vbts,\ext^{n-1}T^{*}\spc}$
as a traction stress. Such a section is represented locally by $(x^{i},\st_{1\dots\wh{\jmath}\dots n\ga}(x^{j}))$,
or specifically, by
\begin{equation}
\sum_{k,r}\st_{1\dots\wh{\jmath}\dots n\ga}g^{\ga}\otimes(\lisuppwout{\dee x}1{\wedge}{\cdots}n{\jmath}).\label{eq:LocalRepOfTractStresses-1}
\end{equation}
The transposed, $\st\trps$, is represented by
\begin{equation}
\sum_{j,\ga}\st_{1\dots\wh{\jmath}\dots n\ga}(\lisuppwout{\dee x}1{\wedge}{\cdots}nj)\otimes g^{\ga}
\end{equation}
and $\st(\vf)$ is represented locally by
\begin{equation}
\sum_{j,\ga}\st_{1\dots\wh{\jmath}\dots n\ga}\vf^{\ga}\lisuppwout{\dee x}1{\wedge}{\cdots}nj.\label{eq:ReprActionOfTrackStOnVelField-1}
\end{equation}
We conclude that in case $\st=p_{\st}(\stm)$, then, 
\begin{equation}
\sigma_{1\dots\wh{\jmath}\dots n\ga}=(-1)^{j-1}\stm_{1\dots n\ga}^{1j}.\label{eq:CauchyAndVariational}
\end{equation}

For each $(n-1)$-dimensional oriented submanifold $\var\subset\spc$,
in particular, the boundary $\bdry\body$ of an $n$-dimensional submanifold
with boundary $\body\subset\spc$, one may integrate $\st(\vf)$ over
$\var$, and evaluate
\begin{equation}
\int_{\var}\incl_{\var}^{*}(\st(\vf)).
\end{equation}
Here, $\incl_{\var}:\var\to\spc$ is the natural inclusion so that
$\incl_{\var}^{*}$ is the restriction of forms. We conclude that
\begin{equation}
\sfc_{\var}=\incl_{\var}^{*}\comp\st\label{eq:Cauchy_Form}
\end{equation}
 is the surface force induced by $\st$ and the integral above represents
the power produced by the traction. The relation (\ref{eq:Cauchy_Form})
is a generalization of the traditional Cauchy formula.

\subsection{The divergence of stress and field equations\label{subsec:The-divergence}}

The divergence, $\diver\stm$, of the variational stress field $\stm$
is a section of $\L{\vbts,\ext^{n}T^{*}\spc}$ which is defined invariantly
by (see \cite{Segev2002,Segev2013})
\begin{equation}
\diver\stm(\vf)=\dee\paren{p_{\st}(S)(\vf)}-\stm(\jet{\vf)},\label{eq:DefineDivergence-1}
\end{equation}
for every differentiable vector field $\vf$. To present the local
expression for $\diver\stm$ we first note that if $\st=p_{\st}(\stm)$,
then $\dee(\st(\vf))$ is represented locally by

\begin{multline}
\sum_{j,\ga}\dee(\st_{1\dots\wh{\jmath}\dots n\ga}\vf^{\ga})\wedge\lisupwout{\dee x}{\wedge}nj\\
\begin{split} & =\sum_{i,j,\ga}(\st_{1\dots\wh{\jmath}\dots n\ga}\vf^{\ga})_{,i}\dee x^{i}\wedge\lisupwout{\dee x}{\wedge}{\ga}j,\\
 & =\sum_{j,\ga}(\st_{1\dots\wh{\jmath}\dots n\ga}\vf^{\ga})_{,j}(-1)^{j-1}\lisup{\dee x}{\wedge}n,\\
 & =\sum_{j,\ga}(S_{1\dots n\ga}^{j}\vf^{\ga})_{,j}\lisup{\dee x}{\wedge}n.
\end{split}
\vphantom{}
\end{multline}
Using Equation (\ref{eq:ReprPowerIntern-1}), the local expression
for $\diver\stm(\vf)$ is therefore
\begin{multline}
\sum_{j,p}\left[(S_{1\dots n\ga}^{j}\vf^{\ga})_{,j}-\left(\sum_{\ga}R_{1\dots n\ga}\vf^{\ga}+\sum_{j,\ga}\stm_{1\dots n\ga}^{j}\vf_{,j}^{\ga}\right)\right]\lisup{\dee x}{\wedge}n\\
=\sum_{j,\ga}(\stm_{1\dots n\ga,j}^{j}-R_{1\dots n\ga})\vf^{\ga}\lisup{\dee x}{\wedge}n
\end{multline}
so that $\diver\stm$ is represented locally by
\begin{equation}
\sum_{j,\ga}(\stm_{1\dots n\ga,j}^{j}-\slf_{1\dots n\ga})g^{\ga}\otimes(\lisup{\dee x}{\wedge}n).\label{eq:ReprDivergence-1}
\end{equation}

It is noted that in the case where $\slf_{1\dots n\ga}=0$ locally,
the expression for the divergence reduces to the traditional expression
for the divergence of a tensor field in a Euclidean space. 

Given a variational stress $\stm$, and setting 
\begin{equation}
\bfc=-\diver\stm,\label{eq:DiffBalanceLawStresses-1}
\end{equation}
 for every $n$-dimensional submanifold with boundary $\body\subset\spc$,
a force $\fc_{\body}$ may be represented in the form 
\begin{equation}
\fc_{B}(\vf)=\int_{\body}\stm(j(\vf))=\int_{\body}\bfc(\vf)+\int_{\bdry\body}\sfc_{\bdry\body}(\vf)\label{eq:PrinVirtWorkSimple}
\end{equation}
which is our generalization of the principle of virtual work.

\section{High Order Stresses\label{sec:High-Order-Stresses}}

For continuum mechanics of order greater than one, the fundamental
object we consider is the $k$-th order variational stress which is
a smooth section of the vector bundle $L(J^{k}\avb,\ext^{n}T^{*}\spc)\isom(J^{k}\avb)^{*}\otimes\ext^{n}T^{*}\spc$,
for some vector bundle $\avb\to\spc$, sections of which are interpreted
as virtual generalized velocities. (See \cite{Segev1986} and \cite{SegevDeBotton1991}
for motivation.) Thus, the virtual power performed by a $k$-th order
variational stress $\stm$ for the virtual generalized velocity $\avf$
in a body $\body\subset\spc$ is given by the action of the functional
\begin{equation}
\fc_{\body}:\avf\longmapsto\int_{\body}\stm(j^{k}\avf).\label{eq:kthOrdrPower}
\end{equation}
Observing (\ref{eq:FiberOfJetBundle}), it follows that the fiber,
$L(J^{k}\avb,\ext^{n}T^{*}\spc)_{x}$ of $L(J^{k}\avb,\ext^{n}T^{*}\spc)$
at $x\in\spc$ is isomorphic with
\begin{equation}
\left(\avb_{x}^{*}\oplus\L{T_{x}\spc,\avb_{x}}^{*}\oplus\cdots\oplus L_{S}^{p}(T_{x}\spc,\avb_{x})^{*}\oplus\cdots\oplus L_{S}^{k}(T_{x}\spc,\avb_{x})^{*}\right)\otimes\ext^{n}T_{x}^{*}\spc,
\end{equation}
where the isomorphism depends on the charts used. Let $\stm^{p}$
denote the component of the representative of $\stm$ in $L_{S}^{p}(T\spc,\vbts)^{*}\otimes\ext^{n}T^{*}\spc$.
It follows that the stress may be represented locally in the form
$(\stm^{0},\stm^{1},\dots,\stm^{k})$, where $\stm^{p}$ is an array
in the form $\stm_{1\dots n\ga}^{p\mii}$, and $\mii$ is a multi-index
with $\abs{\mii}=p$. The action $\stm(A)$ for an element $A\in J^{k}\avb$
is given by
\begin{equation}
\sum_{p=\abs{\mii}\les k,\,\ga}\stm_{1\dots n\ga}^{p\mii}A_{\mii}^{p\ga}\dee x^{1}\wedge\cdots\wedge\dee x^{n}.
\end{equation}
For $A=j^{k}\avf$,
\begin{equation}
\stm(A)=\sum_{p=\abs{\mii}\les k,\,\ga}\stm_{1\dots n\ga}^{p\mii}\avf_{,\mii}^{\ga}\dee x^{1}\wedge\cdots\wedge\dee x^{n}.
\end{equation}
Explicitly, $\stm$ is given locally in the form
\begin{equation}
\sum\left(S_{1\dots n\ga}^{0}+\stm_{1\dots n\ga}^{1i_{1}}\partial_{i_{1}}+\cdots+\stm_{1\dots n\ga}^{ki_{1}\dots i_{k}}\partial_{i_{1}}\otimes\cdots\otimes\partial_{i_{k}}\right)\otimes g^{\ga}\otimes\dee x^{1}\wedge\cdots\wedge\dee x^{n},\label{eq:RepHStress}
\end{equation}
and the action is 
\begin{equation}
\stm(A)=\sum\stm_{1\dots n\ga}^{pi_{1}\dots i_{p}}\avf_{,i_{1}\dots i_{p}}^{\ga}\dee x^{1}\wedge\cdots\wedge\dee x^{n},
\end{equation}
where the sums are taken over all $i_{1},\dots,i_{k}=1,\dots,n$,
and $p=0,\dots,k$. Evidently, the arrays $\stm_{1\dots n\ga}^{pi_{1}\dots i_{p}}$
are symmetric in all $i$ indices.

It is our objective to represent the virtual power for high order
stresses (\ref{eq:kthOrdrPower}) in a form analogous to (\ref{eq:PrinVirtWorkSimple}).
.

\subsection{Significant components of hyper-stresses\label{subsec:Significant-components-HS}}

The inclusion $\incl_{V}^{r}$ (\ref{eq:Incl_Verts}) of the vertical
subbundles induces a projection
\begin{equation}
\incl_{V}^{r*}:L(J^{k}\avb,\ext^{n}T^{*}\spc)\tto L(V^{r}J^{k}\avb,\ext^{n}T^{*}\spc),
\end{equation}
by $\incl_{V}^{r*}(\stm):=\stm\comp\incl_{V}^{r}$. Evidently, the
representatives of $\incl_{V}^{r*}(\stm)$ depend only on $\stm_{1\dots n\ga}^{pI}$
for $p>r$. In fact, $L(V^{r}J^{k}\avb,\ext^{n}T^{*}\spc)\resto x$
is isomorphic with
\begin{equation}
\left(L_{S}^{r+1}(T_{x}\spc,\avb_{x})^{*}\oplus\cdots\oplus L_{S}^{k}(T_{x}\spc,\avb_{x})^{*}\right)\otimes\ext^{n}T_{x}^{*}\spc.\label{eq:Rep_Dir_Sum}
\end{equation}
Specifically, for the representation of a hyper-stress $\stm$ as
in (\ref{eq:RepHStress}), $\incl_{V}^{r*}(\stm)$ is 
\begin{multline}
\sum\left(\stm_{1\dots n\ga}^{r+1i_{1}\dots i_{r+1}}\partial_{i_{1}}\otimes\cdots\otimes\partial_{i_{r+1}}+\cdots+\right.\\
\left.\stm_{1\dots n\ga}^{ki_{1}\dots i_{k}}\partial_{i_{1}}\otimes\cdots\otimes\partial_{i_{k}}\right)\otimes g^{\ga}\otimes\dee x^{1}\wedge\cdots\wedge\dee x^{n}.\label{eq:Rep_r_VertHyp}
\end{multline}

In particular, for the case $r=k-1$, one has a natural isomorphism
\begin{equation}
L(V^{k-1}J^{i}\avb,\ext^{n}T^{*}\spc)\isom L_{S}^{k}(T\spc,\avb)^{*}\otimes\ext^{n}T_{x}^{*}\spc
\end{equation}
and a natural 
\begin{equation}
\incl_{V}^{k-1*}:L(J^{k}\avb,\ext^{n}T^{*}\spc)\tto L_{S}^{k}(T\spc,\avb)^{*}\otimes\ext^{n}T_{x}^{*}\spc.
\end{equation}
which isolates the significant high-order components of $k$-order
stresses. When stresses are viewed as linear differential operators,
the significant components are the symbols of the differential operators
in the terminology of \cite{Palais68}.

For the particular case $r=0$, one has, locally,
\begin{equation}
V^{0}J^{k}\avb\resto x\isom L_{S}^{1}(T_{x}\spc,\avb_{x})\oplus\cdots\oplus L_{S}^{k}(T_{x}\spc,\avb_{x})
\end{equation}
and 
\begin{equation}
\incl_{V}^{0*}:L(J^{k}\avb,\ext^{n}T^{*}\spc)\tto L(V^{0}J^{k}U,\ext^{n}T^{*}\spc).
\end{equation}
Thus, $\incl_{V}^{0*}(\stm)$ is represented locally by an element
of $[L^{1}(T\spc,\avb)^{*}\oplus\cdots\oplus L_{S}^{k}(T\spc,\avb)^{*}]\otimes\ext^{n}T^{*}\spc$.

\begin{rem}
In order to continue the reduction process in analogy with Section
\ref{sec:Simple-Stresses}, one may consider performing a contraction
operator in analogy with Equation (\ref{eq:contraction-map}) so that
$\spec C(\incl_{V}^{r*}(\stm))$ be represented locally by
\begin{multline}
\sum\left(\stm_{1\dots n\ga}^{r+1i_{1}\dots i_{r+1}}\partial_{i_{2}}\otimes\cdots\otimes\partial_{i_{r+1}}+\cdots+\right.\\
\left.\stm_{1\dots n\ga}^{ki_{1}\dots i_{k}}\partial_{i_{2}}\otimes\cdots\otimes\partial_{i_{k}}\right)\otimes g^{\ga}\otimes\left(\partial_{i_{1}}\contr\dee x^{1}\wedge\cdots\wedge\dee x^{n}\right).\label{eq:Rep_r_VertHyp-1-1}
\end{multline}
However, the simplest case of $L(V^{0}J^{2}\avb,\ext^{n}T^{*}\spc)$
may serve as a counter-example. Let $S\in L(V^{0}J^{2}\avb,\ext^{n}T^{*}\spc)$
be represented in two charts $(x^{i},\avf^{\ga})$ and $(x^{i'},\avf^{\ga'})$
so that $u^{\ga'}=A_{\ga}^{\ga'}\avf^{\ga}$, in the forms
\begin{equation}
\begin{split}\stm & =\sum_{i,j,\ga}(S_{1\dots n\ga}^{1i}\partial_{i}+\stm_{1\dots n\ga}^{2ij}\partial_{i}\otimes\bdry_{j})\otimes g^{\ga}\otimes(\dee x^{1}\wedge\cdots\wedge\dee x^{n})\\
 & =\sum_{i',j',\ga'}(S_{1'\dots n'\ga'}^{1i'}\partial_{i'}+\stm_{1'\dots n'\ga'}^{2i'j'}\partial_{i'}\otimes\bdry_{j'})\otimes g^{\ga'}\otimes(\dee x^{1'}\wedge\cdots\wedge\dee x^{n'}).
\end{split}
\end{equation}
Thus, one might consider defining $\spec C(\stm)$ locally by 
\begin{equation}
\sum_{i',j',\ga'}(S_{1'\dots n'\ga'}^{1i'}+\stm_{1'\dots n'\ga'}^{2i'j'}\bdry_{j'})\otimes g^{\ga'}\otimes(\partial_{i'}\contr(\dee x^{1'}\wedge\cdots\wedge\dee x^{n'}))\label{eq:ContrAtmpPrime}
\end{equation}
with an analogous expression for the representation in terms of $(x^{i},\avf^{\ga})$.

We have the following transformations.
\begin{gather}
\avf_{,i'}^{\ga'}=\sum_{\ga,i}(A_{\ga,i}^{\ga'}x_{i'}^{i}\avf^{\ga}+A_{\ga}^{\ga'}\avf_{,i}^{\ga}x_{,i'}^{i}),
\end{gather}
\begin{multline}
\avf_{,i'j'}^{\ga'}=\sum_{\ga,i,j}\left[(A_{\ga,ij}^{\ga'}x_{,i'}^{i}x_{,j'}^{j}+A_{\ga,i}^{\ga'}x_{,i'j'}^{i})\avf^{\ga}+\right.\\
\left.(2A_{\ga,j}^{\ga'}x_{,i'}^{j}x_{,j'}^{i}+A_{\ga}^{\ga'}x_{,i'j'}^{i})\avf_{,i}^{\ga}+A_{\ga}^{\ga'}x_{,i'}^{i}x_{,j'}^{j}\avf_{,ij}^{\ga}\right].
\end{multline}
Comparing the local expressions for $\stm(j^{2}\avf(x))$ for a section
$\avf$ with $\avf(x)=0$, and using 
\begin{equation}
\dee x^{1'}\wedge\cdots\wedge\dee x^{n'}=J\dee x^{1}\wedge\cdots\wedge\dee x^{n},\quad J=\det(x_{,i}^{i'}),
\end{equation}
one obtains
\begin{equation}
\stm_{1\dots n\ga}^{1i}=J\sum_{\ga',i',j'}\left[\stm_{1'\dots n'\ga'}^{1i'}A_{\ga}^{\ga'}x_{,i'}^{i}+S_{1'\dots n'\ga'}^{2i'j'}(2A_{\ga,j}^{\ga'}x_{,i'}^{j}x_{,j'}^{i}+A_{\ga}^{\ga'}x_{,i'j'}^{i})\right],
\end{equation}
\begin{equation}
\stm_{1\dots n\ga}^{2ij}=J\sum_{\ga',i',j'}S_{1'\dots n'\ga'}^{2i'j'}A_{\ga}^{\ga'}x_{,i'}^{i}x_{j'}^{j}.
\end{equation}
Substituting these relations into the analog of Equation (\ref{eq:ContrAtmpPrime})
gives
\begin{multline}
\sum_{i,j,\ga}(S_{1\dots n\ga}^{1i}+\stm_{1\dots n\ga}^{2ij}\bdry_{j})\otimes g^{\ga}\otimes(\partial_{i}\contr(\dee x^{1}\wedge\cdots\wedge\dee x^{n}))\\
=J\sum_{\ga',i',j'}\left[\stm_{1'\dots n'\ga'}^{1i'}A_{\ga}^{\ga'}x_{,i'}^{i}+S_{1'\dots n'\ga'}^{2i'j'}(2A_{\ga,j}^{\ga'}x_{,i'}^{j}x_{,j'}^{i}+A_{\ga}^{\ga'}x_{,i'j'}^{i})\right.\\
+\left.S_{1'\dots n'\ga'}^{2i'j'}A_{\ga}^{\ga'}x_{,i'}^{i}x_{j'}^{j}x_{,j}^{k'}\bdry_{k'}\right]\otimes A_{\gb'}^{\ga}g^{\gb'}\\
\otimes(x_{,i}^{l'}\bdry_{l'}\contr(\dee x^{1'}\wedge\cdots\wedge\dee x^{n'})/J).
\end{multline}
The last expression is not of the form (\ref{eq:ContrAtmpPrime})
because of the second term in the sum. We conclude, therefore, that
the proposed definition of the contraction is not invariant. In fact,
had the contraction been invariant, this would imply an invariant
decomposition $\stm\mapsto(S^{1},S^{2})$.
\end{rem}

\begin{rem}
\label{rem:Contraction}One could also consider applying the contraction
to an element $\stm$ of $L(V^{k-1}J^{k}\avb,\ext^{n}T^{*}\spc)=L(L_{S}^{k}(T\spc,\avb),\ext^{n}T^{*}\spc)$.
Thus, for a representation 
\begin{equation}
\stm=\sum_{\ga,i_{1},\dots,i_{k}}S_{1\cdots n\ga}^{ki_{1}\dots i_{k}}\bdry_{i_{1}}\otimes\cdots\otimes\bdry_{i_{k}}\otimes g^{\ga}\otimes(\dee x^{1}\wedge\cdots\wedge\dee x^{n}),
\end{equation}
one may set $\spec C(\stm)\in L(L_{S}^{k-1}(T\spc,\avb),\ext^{n-1}T^{*}\spc)$
by
\begin{equation}
\spec C(\stm)=\sum_{\ga,i_{1},\dots,i_{k}}S_{1\cdots n\ga}^{ki_{1}\dots i_{k}}\bdry_{i_{2}}\otimes\cdots\otimes\bdry_{i_{k}}\otimes g^{\ga}\otimes(\bdry_{i_{1}}\contr(\dee x^{1}\wedge\cdots\wedge\dee x^{n})).
\end{equation}
 While this definition is invariant, it does not lead to any meaningful
result because we do not have an object in $L_{S}^{k-1}(T\spc,\avb)$
to apply $\spec C(\stm)$ to. In particular, it is observed that the
only section of $\avb$ whose jet is $r$-vertical, for any $r$,
is the zero section. As we will show below, further contractions vanish
due to the symmetry of $S_{1\cdots n\ga}^{ki_{1}\dots i_{k}}$ relative
to the $i$-indices.

Another unsuccessful attempt would be to identify elements of $V^{p}J^{k}\avb$
as elements of $J^{k-p-1}(L_{S}^{p+1}(T\spc,\avb))$. Again, a local
expression may be shown to be non-invariant. 

A way to overcome these difficulties will be suggested below by embedding
the space of $k$-jets in the space of iterated jets.
\end{rem}

\section{Iterated Jet Bundles\label{sec:Iterated-Jet-Bundle}}

The method proposed below for manipulating hyper-stresses uses the
representation of hyperstresses by non-holonomic hyper-stresses, which
are defined on iterated jet bundles. This section describes the basic
definitions associated with iterated jet bundles.

\subsection{The iterated jet bundle $J^{1}(J^{1}\protect\avb)$.}

Since for any vector bundle $\avb$, $\vbts=J^{1}\avb\to\spc$ is
also a vector bundle, one may consider the vector bundle $J^{1}\vbts=J^{1}(J^{1}U)\to\spc$.
In particular, using the subscript $J^{1}\avb$ to indicate that $J^{1}\avb$
is the vector bundle to which the jet projections correspond, we have
projections 
\begin{equation}
\pi_{0J^{1}\avb}^{1}:J^{1}(J^{1}\avb)\tto J^{1}\avb,\qquad\pi_{J^{1}\avb}^{1}:J^{1}(J^{1}\avb)\tto\spc.
\end{equation}
As any section $A$ of $J^{1}\avb$, or of $\pi_{\avb}^{1}$ to be
specific, is locally of the form 
\begin{equation}
x\longmapsto(\avf^{\ga}(x),A_{i}^{\gb}(x))=(A^{0\ga_{0}}(x),A_{i}^{1\ga_{1}}(x)),
\end{equation}
an element $B$ of the iterated jet bundle at the point $x\in\spc$
is of the form 
\[
(x^{i},B^{0\ga_{0}},B_{i_{1}}^{1\ga_{1}},B_{i_{2}}^{2\ga_{2}},B_{i_{3}i_{4}}^{3\ga_{3}}).
\]
Here, $(B^{0\ga_{0}},B_{i_{1}}^{1\ga_{1}})$ represent $B_{0}=\pi_{0J^{1}\avb}^{1}(B)$\textemdash the
value of a section $A$ of $J^{1}\avb$ at $x$, and $(B_{i_{2}}^{2\ga_{2}},B_{i_{3}i_{4}}^{3\ga_{3}})$
represent the derivative of the section $A$ at $x$, \ie,  
\[
(B_{i_{2}}^{2\ga_{2}},B_{i_{3}i_{4}}^{3\ga_{3}})=(A_{,i_{2}}^{0\ga_{2}},A_{i_{3},i_{4}}^{1\ga_{3}}).
\]

It is noted that there is a natural vector bundle inclusion $\incl:J^{2}\avb\to J^{1}(J^{1}\avb)$
such that $\incl_{x}:J^{2}\avb_{x}\to J^{1}(J^{1}\avb)_{x}$ is given
as follows. Let $\avf$ be a section of $\avb$ that represents an
element $B\in J^{2}\avb_{x}$. Then, $\vf=j^{1}\avf$ is a section
of $J^{1}\avb$ whose jet is the target element $\incl(B)=j^{1}\vf(x)=j^{1}(j^{1}\avf)(x)$
in $J^{1}(J^{1}\avb)$. Thus, locally,
\begin{equation}
(\avf^{\ga_{0}},\avf_{,i_{1}}^{\ga_{1}},\avf_{,i_{1}i_{2}}^{\ga_{2}})\longmapsto(\avf^{\ga_{0}},\avf_{,i_{1}}^{\ga_{1}},\avf_{,i_{2}}^{\ga_{2}},\avf_{,i_{3}i_{4}}^{\ga_{3}}).
\end{equation}
Evidently, the result is independent of the section chosen and locally
the inclusion is in the form
\begin{equation}
(A^{0\ga_{0}},A_{i_{1}}^{1\ga_{1}},A_{i_{1}i_{2}}^{2\ga_{2}})\longmapsto(A^{0\ga_{0}},A_{i_{1}}^{1\ga_{1}},A_{i_{2}}^{1\ga_{2}},A_{i_{3}i_{4}}^{2\ga_{3}}).
\end{equation}
Thus, the image of $\incl$ contains elements for which the second
and third groups of components are identical and the components in
the fourth group are symmetric.

It is observed finally (see \cite[p.~169]{Saunders}) that there is
no natural inverse to $\incl$, \ie, a projection $J^{1}(J^{1}\avb)\to J^{2}\avb$.

An additional projection 
\begin{equation}
j^{1}\pi_{0\avb}^{1}:J^{1}(J^{1}\avb)\tto J^{1}U,
\end{equation}
may be defined as follows. Consider the lift (see (\ref{eq:Lift_of_VBM}))
$j^{1}\pi_{0U}^{1}:J^{1}(J^{1}\avb)\to J^{1}\avb$. Then,  if $B\in J^{1}(J^{1}\avb)\resto x$
is represented by a section $A$ of $J^{1}\avb$, $j^{1}\pi_{0\avb}^{1}(A)$
is given by $j^{1}\pi_{0U}^{1}(B)=j^{1}(\pi_{0\avb}^{1}\comp A)(x)$.
Locally, $J^{1}\pi_{0\avb}^{1}$ is represented by 
\begin{equation}
(x^{i},B^{0\ga_{0}},B_{i_{1}}^{1\ga_{1}},B_{i_{2}}^{2\ga_{2}},B_{i_{3}i_{4}}^{3\ga_{3}})\longmapsto(x^{i},B^{0\ga_{0}},B_{i_{2}}^{2\ga_{2}}).
\end{equation}

Various levels of ``compatibility'' or ``holonomicity'' may be considered
for sections of the iterated jet bundle. Let $B$ be a section of
$\pi_{J^{1}\avb}^{1}$ and $B_{0}=\pi_{0J^{1}U}^{1}\comp B$. The
simplest condition requires that $B_{0}$ is compatible, that is,
$B_{0}=j^{1}u$ for some section $\avf$ of $\avb$. It follows that
$B_{0j}^{\gb}=u_{,j}^{\gb}$ and that \foreignlanguage{english}{$\pi_{0J^{1}\avb}^{1}\comp B=j^{1}(\pi_{0U}^{1}\comp\pi_{0J^{1}\avb}^{1}\comp B)$}.
Another compatibility condition is that $B=j^{1}A_{0}$, where $A_{0}$
is a section of $J^{1}\avb$. In such a case, $j^{1}(\pi_{0J^{1}\avb}^{1}\comp B)=B$,
and if $(B^{0\ga_{0}},B_{i_{1}}^{1\ga_{1}})$ represent $A_{0}$,
then, $B_{i}^{2\ga}=B_{,i}^{0\ga}$ and $B_{jk}^{3\ga}=B_{j,k}^{1\ga}.$
If these conditions hold, one refers to $B$ as semi-holonomic. A
section $\avf$ of $\avb$ induces a holonomic section $B$ of $J^{1}(J^{1}\avb)$
by setting $A_{0}=j^{1}\avf$, $B=j^{1}A_{0}=j^{1}(j^{1}\avf)$. In
other words, $j^{1}(j^{1}(\pi_{0\avb}^{1}\comp\pi_{0J^{1}\avb}^{1}\comp B))=B$
and it is observed that $j^{1}(j^{1}\pi_{0\avb}^{1}\comp B)=j^{1}(\pi_{0J^{1}\avb}^{1}\comp B)$.
 Note that if no holonomicity is imposed, $B_{i_{3}i_{4}}^{3\ga_{3}}$
need not be symmetric in the $i_{3},\,i_{4}$ indices.

Next, we consider vertical subbundles whose sections are non-holonomic.
The projection $\pi_{0J^{1}\avb}^{1}$ determines the subbundle
\begin{equation}
V_{23}J^{1}(J^{1}\avb):=\kernel\pi_{0J^{1}\avb}^{1}
\end{equation}
whose elements are of the form $(x^{i},B^{0\ga_{0}}=0,B_{i_{1}}^{1\ga_{1}}=0,B_{i_{2}}^{2\ga_{2}},B_{i_{3}i_{4}}^{3\ga_{3}})$.
We will use $\incl_{23}$ to denote its inclusion in $J^{1}(J^{1}\avb)$.
In addition, we have the vector subbundle $V_{13}J^{1}(J^{1}\avb)=\kernel j^{1}\pi_{0\avb}^{1}$
whose elements are of the form $(x^{i},B^{0\ga_{0}}=0,B_{i_{1}}^{1\ga_{1}},B_{i_{2}}^{2\ga_{2}}=0,B_{i_{3}i_{4}}^{3\ga_{3}})$.
One can easily confirm that 
\begin{equation}
V_{13}J^{1}(J^{1}\avb)=\kernel j^{1}\pi_{0\avb}^{1}\isom J^{1}(VJ^{1}\avb),
\end{equation}
where $VJ^{1}\avb=\kernel\pi_{0\avb}^{1}$. We will denote its inclusion
in $J^{1}(J^{1}\avb)$ by $\incl_{13}$.

As a result, one has the vector subbundle of completely vertical iterated
jets 
\begin{equation}
V_{3}J^{1}(J^{1}\avb):=\kernel\pi_{0J^{1}\avb}^{1}\cap\kernel j^{1}\pi_{0\avb}^{1}
\end{equation}
whose elements are represented in the form $(x^{i},B^{0\ga_{0}}=0,B_{i_{1}}^{1\ga_{1}}=0,B_{i_{2}}^{2\ga_{2}}=0,B_{i_{3}i_{4}}^{3\ga_{3}})$.
We will use $\incl_{3}$ to denote its inclusion in $J^{1}(J^{1}\avb)$.
Evidently,
\begin{equation}
V_{3}J^{1}(J^{1}\avb)\isom L^{2}(T\spc,\avb).
\end{equation}
We conclude that there is a natural inclusion 
\begin{equation}
\incl_{S}:V^{1}J^{2}\avb\isom L_{S}^{2}(T\spc,\avb)\hookrightarrow V_{3}J^{1}(J^{1}\avb)\isom L^{2}(T\spc,\avb)
\end{equation}
of completely vertical 2-jets in the vector subbundle of completely
vertical iterated jets. In addition, there is a natural projection
induced by symmetrization

\begin{equation}
\pi_{S}:V_{3}J^{1}(J^{1}\avb)\isom L^{2}(T\spc,\avb)\tto V^{1}J^{2}\avb\isom L_{S}^{2}(T_{x}\spc,\avb_{x}),\quad B\longmapsto\shalf(B+B^{T}).
\end{equation}

Finally, the vector subbundle $V_{123}J^{1}(J^{1}\avb)=\kernel(\pi_{0\avb}^{1}\comp\pi_{0J^{1}\avb}^{1})$
contains elements represented in the form $(x^{i},B^{0\ga_{0}}=0,B_{i_{1}}^{1\ga_{1}},B_{i_{2}}^{2\ga_{2}},B_{i_{3}i_{4}}^{3\ga_{3}})$.
The inclusion $V_{123}J^{1}(J^{1}\avb)\hookrightarrow J^{1}(J^{1}\avb)$
will be denoted in analogy by $\incl_{123}$.

\subsection{The iterated jet bundle $J^{p}(J^{r}\protect\avb)$.}

In more general situations, one may consider the iterated jet bundle
$J^{p}(J^{r}\avb)$. Again, one has an inclusion 
\begin{equation}
\incl_{p,r}^{p+r}:J^{p+r}\avb\hookrightarrow J^{p}(J^{r}\avb)\label{eq:Inclusion_r-Vert}
\end{equation}
 given as follows. Let $\avf$ be a section that represents an element
of $(J^{p+r}\avb)_{x}$. Then, $J^{r}\avf$ is a section of the vector
bundle $J^{r}\avb\to\spc$, and so, $j^{p}(j^{r}\avf)(x)\in(J^{p}(J^{r}\avb))_{x}$.

We also observe that the $q$-lift of this inclusion is
\begin{equation}
j^{q}\incl_{p,r}^{p+r}:J^{q}(J^{p+r}\avb)\tto J^{q}(J^{p}(J^{r}\avb)).
\end{equation}
Thus, using the inclusion $\incl_{q,p+r}^{p+r+q}:J^{p+r+q}\avb\to J^{q}(J^{p+r}\avb)$,
gives the inclusions
\begin{equation}
j^{q}\incl_{p,r}^{p+r}\comp\incl_{q,p+r}^{p+r+q}:J^{p+r+q}\avb\hookrightarrow J^{q}(J^{p}(J^{r}\avb)).\label{eq:Iterated_inclusion}
\end{equation}

Evidently, one can continue inductively and include any jet bundle
in a multiply-iterative jet bundle. For a vector bundle $\avb$, we
will use the notation 
\begin{equation}
\incl_{\mathrm{it}}:J^{k}\avb\tto(J^{1})^{k}\avb:=\underbrace{J^{1}(J^{1}(\cdots\cdots J^{1}(J^{1}\avb)\cdots))}_{k\text{-times}}\label{eq:Include_in_k_Iterated}
\end{equation}
for the vector bundle injection of the $k$-th jet bundle in the $k$-times
iterated $(J^{1})^{k}\avb$ jet bundle of $\avb$.

\section{Non-Holonomic Hyper-Stresses\label{sec:Non-Holonomic-Stresses}}

Just as elements of $L(J^{2}\avb,\ext^{n}T^{*}\spc)$ represent second
order hyper-stresses, we refer to elements of $L(J^{1}(J^{1}\avb),\ext^{n}T^{*}\spc)$
as non-holonomic hyper-stresses. The inclusion $\incl:J^{2}\avb\to J^{1}(J^{1}\avb)$
induces a projection 
\begin{equation}
\incl^{*}:L(J^{1}(J^{1}\avb),\ext^{n}T^{*}\spc)\tto L(J^{2}\avb,\ext^{n}T^{*}\spc),\qquad X\longmapsto X\comp\incl.
\end{equation}
Thus, every $2$-hyper-stress may be represented by a non-holonomic
hyper-stress.

This argument applies also to the inclusion of the $k$-times iterated
jet bundle and so one has a representation
\begin{equation}
\incl_{\mathrm{it}}^{*}:L((J^{1})^{k}\avb,\ext^{n}T^{*}\spc)\tto L(J^{k}\avb,\ext^{n}T^{*}\spc).
\end{equation}
For this reason, we will continue the analysis for $2$-hyper-stresses
only. The general case follows inductively.

\subsection{Basic properties}

A non-holonomic hyper-stress field may be represented in the form
\begin{equation}
(x^{i},X_{1\dots n\ga_{0}}^{0},X_{1\dots n\ga_{1}}^{1i_{1}},X_{1\dots n\ga_{2}}^{2i_{2}},X_{1\dots n\ga_{3}}^{3i_{3}i_{4}}),
\end{equation}
or more explicitly, in the form
\begin{multline}
\left(\lsum_{\ga}X_{1\dots n\ga}^{0}g^{\ga},\lsum_{\ga,i}X_{1\dots n\ga}^{1i}\partial_{i}\right.\otimes g^{\ga},\lsum_{\ga,i}X_{1\dots n\ga}^{2i}\partial_{i}\otimes g^{\ga},\\
\lsum_{\ga,i,j}X_{1\dots n\ga}^{3ij}\partial_{i}\otimes\left.\partial_{j}\otimes g^{\ga}\right)\otimes(\lisupc{\dee x}{\wedge}n),\label{eq:rep-nhhs}
\end{multline}
where we have omitted the indication of the dependence of the various
fields on $x\in\spc$.

With the notation introduced above, the action of a non-holonomic
hyper-stress on an element $B$ of $J^{1}(J^{1}\avb)$ is given as
\begin{equation}
\sum_{\ga,i,j}\left(X_{1\dots n\ga}^{0}B^{0\ga}+X_{1\dots n\ga}^{1i}B_{i}^{1\ga}+X_{1\dots n\ga}^{2i}B_{i}^{2\ga}+X_{1\dots n\ga}^{3ij}B_{ij}^{3\ga}\right)\lisupc{\dee x}{\wedge}n.
\end{equation}
For the case where $B=j^{1}A$, for a section $A$ of $J^{1}\avb$,
the expression is
\begin{equation}
\sum_{\ga,i,j}\left(X_{1\dots n\ga}^{0}A^{0\ga}+X_{1\dots n\ga}^{1i}A_{i}^{1\ga}+X_{1\dots n\ga}^{2i}A_{,i}^{0\ga}+X_{1\dots n\ga}^{3ji}A_{j,i}^{1\ga}\right)\lisupc{\dee x}{\wedge}n.
\end{equation}

For an element $A\in J^{2}\avb$, one has
\begin{equation}
\begin{split}\incl^{*}(X)(A) & =X(\incl A)\\
 & =\left(\sum_{\ga}X_{1\dots n\ga}^{0}A^{0\ga}+\sum_{\ga,i}(X_{1\dots n\ga}^{1i}+X_{1\dots n\ga}^{2i})A_{i}^{1\ga}\right.\\
 & \qquad\qquad\qquad\left.+\sum_{\ga}X_{1\dots n\ga}^{3ij}A_{ij}^{2\ga}\right)\lisupc{\dee x}{\wedge}n,\\
 & =\left(\sum_{\ga}X_{1\dots n\ga}^{0}A^{0\ga}+\sum_{\ga,i}(X_{1\dots n\ga}^{1i}+X_{\ga}^{2i})A_{i}^{1\ga}\right..\\
 & \qquad\qquad\left.+\sum_{\ga}\shalf(X_{1\dots n\ga}^{3ij}+X_{1\dots n\ga}^{3ji})A_{ij}^{2\ga}\right)\lisupc{\dee x}{\wedge}n
\end{split}
\end{equation}
The mapping $\incl^{*}$ is therefore a restriction represented locally
by 
\begin{multline}
(x^{i},X_{1\dots n\ga_{0}}^{0},X_{1\dots n\ga_{1}}^{1i_{1}},X_{1\dots n\ga_{2}}^{2i_{2}},X_{1\dots n\ga_{3}}^{3i_{3}i_{4}})\longmapsto\\
(x^{i},X_{1\dots n\ga_{0}}^{0},X_{1\dots n\ga_{1}}^{1i_{1}}+X_{1\dots n\ga_{1}}^{2i_{1}},\shalf(X_{1\dots n\ga_{3}}^{3i_{3}i_{4}}+X_{1\dots n\ga_{3}}^{3i_{4}i_{3}}))
\end{multline}
and being surjective, every second order stress $\stm$ is of the
form 
\begin{equation}
\stm=\incl^{*}X\label{eq:stressInducedByNHS}
\end{equation}
for some non-unique section $X$ of $L(J^{1}(J^{1}U),\ext^{n}T^{*}\spc)$.
Thus, whatever properties we deduce for elements of $J^{1}(J^{1}\avb)$
will hold for their restrictions to $\image\incl$. 

We focus our attention in this section to the analysis of the action
of a non-holonomic hyper-stress field $X$ on a section $A$ of $J^{1}\avb$
in the form
\begin{equation}
A\longmapsto\int_{\body}X(j^{1}(A)),\label{eq:ActionNHSonJet}
\end{equation}
and in particular, the compatible case where $A=j^{1}\avf$, for a
section $\avf$ of $\avb$.

Using the operator $p_{\st}$ as in (\ref{eq:ProjectionToTractionSt}),
one can extract a section $Y=p_{\st}X$ of the vector bundle $L(J^{1}\avb,\ext^{n-1}T^{*}\spc)$.
The local representation of $Y$ is 
\begin{equation}
\sum_{j,i,\ga}\left(Y_{1\dots\wh{\jmath}\dots n\ga}^{0}+Y_{1\dots\wh{\jmath}\dots n\ga}^{1i}\partial_{i}\right)\otimes g^{\ga}\otimes(\lisuppwout{\dee x}1{\wedge}{\cdots}nj)\label{eq:Rep-PsigmaX}
\end{equation}
From (\ref{eq:CauchyAndVariational}) it follows that 
\begin{equation}
Y_{1\dots\wh{\jmath}\dots n\ga}^{0}=(-1)^{j-1}X_{1\dots n\ga}^{2j},\qquad Y_{1\dots\wh{\jmath}\dots n\ga}^{1i}=(-1)^{j-1}X_{1\dots n\ga}^{3ij}.\label{eq:Rep-PsigmaX-1}
\end{equation}

We may now apply the general definitions of the operators $p_{\st}$
and $\mathrm{div}$ to the case of non-holonomic hyper-stresses. Equation
(\ref{eq:DefineDivergence-1}) assumes the form
\begin{equation}
\diver X(A)=\dee\paren{p_{\st}(X)(A)}-X(j^{1}A),
\end{equation}
in which $\diver X$, a section of $\L{J^{1}\avb,\ext^{n}(T^{*}\spc)}$
is represented locally by
\begin{equation}
\sum_{\ga,i,j}\left(X_{1\dots n\ga,j}^{2j}-X_{1\dots n\ga}^{0}+(X_{1\dots n\ga,j}^{3ij}-X_{1\dots n\ga}^{1i})\partial_{i}\right)\otimes g^{\ga}\otimes(\lisup{\dee x}{\wedge}n).\label{eq:repDivX}
\end{equation}

One conclude that for $Y=p_{\st}(X)$, 
\begin{equation}
\int_{\body}X(j^{1}A)=\int_{\body}\dee(Y(A))-\int_{\body}\diver X(A)=\int_{\bdry\body}Y(A)-\int_{\body}\diver X(A).\label{eq:firstIntegrationByParts}
\end{equation}
Our attention will be focused henceforth on the boundary integral.

\subsection{The hyper-surface stress}

We will refer to $Y=p_{\st}(X)$ as a hyper-surface stress. The integration
of $Y(j^{1}\vf)$ on the boundary $\bdry\body$ as above induce the
special effects associated with hyper-stresses, such as surface tension
and edge interaction. Thus, most of the material below is concerned
with the analysis of this term.

Let $\var$ be an $(n-1)$-dimensional submanifold of $\spc$ and
let $\rho_{\var}$ be the restriction of $(n-1)$ forms defined on
$\spc$ to forms on $\var$ which is induced by the inclusion $T\var\to T\spc$.
Thus, for a given $Y\in L(J^{1}\avb,\ext^{n-1}T^{*}\spc)$ and a submanifold
$\var$ $\rho_{\var}\comp Y\in L((J^{1}U)\resto{\var},\ext^{n-1}T^{*}\var)$.

As in the case of stresses, the inclusion$\incl_{V}^{0}:V^{0}J^{1}\avb\tto J^{1}\avb.$
induces 
\begin{equation}
\incl_{V}^{0*}:L(J^{1}\avb,\ext^{n-1}T^{*}\spc)\tto L(L(T\spc,U),\ext^{n-1}T^{*}\spc),\qquad Y\lmt Y\comp\incl_{V}^{0},
\end{equation}
represented locally by
\begin{equation}
(x,Y_{1\dots\wh{\imath}\dots n\ga}^{0},Y_{1\dots\wh{\jmath}\dots n\ga}^{1})\lmt(x,Y_{1\dots\wh{\jmath}\dots n\ga}^{1}),
\end{equation}
where a basis $\{\lisupwout{\dee x}{\wedge}ni\}$ of $\ext^{n-1}T^{*}\spc$
has been used.

Let $(y^{a})$, $a=1,\dots,n-1$ be local coordinates in an open set
of $\var$ so that $(y^{1},\dots,y^{n-1},x^{n})$ are local coordinates
in $\spc$. The local representative of $\rho_{\var}\comp Y$ is of
the form 
\begin{equation}
\sum_{a,\ga}\left(Y_{1\dots n-1\ga}^{0}+\left(Y_{1\dots n-1\ga}^{1a}\partial_{a}+Y_{1\dots n-1\ga}^{1n}\partial_{n}\right)\right)\otimes g^{\ga}\otimes(\lisuppc{\dee y}1{\wedge}{n-1})
\end{equation}
and that of $\incl_{V}^{0*}(Y)$ is
\begin{equation}
\sum_{a,\ga}\left(Y_{1\dots n-1\ga}^{1a}\partial_{a}+Y_{1\dots n-1\ga}^{1n}\partial_{n}\right)\otimes g^{\ga}\otimes(\lisuppc{\dee y}1{\wedge}{n-1}).
\end{equation}
Here, $\partial_{a}$ denotes the tangent basis vector $\partial/\partial y^{a}$. 
\begin{rem}
\label{rem:Relation_To_H-stress}It is observed that since the bundle
$\avb$ under consideration is a general vector bundle, it may be,
in particular, the jet bundle of any other vector bundle, and so,
the procedure above may be used in an iterative manner for hyper-stresses
of any order.

In particular, consider the case where $\avb=J^{k-2}\vbts$, for a
vector bundle $\vbts$. The local representation of a section of $\avb$
is of the form 
\[
(x,\vf^{\ga},A_{i_{1_{1}}}^{1\ga},\dots,A_{i_{(k-2)_{1}}\dots i_{(k-2)_{k-2}}}^{(k-2)\ga})
\]
 where each $i$-index has two identifiers the first of which indicates
the order of the corresponding tensor in accordance with the superscript.
Thus, an element of $J^{1}U$ is represented locally in of the form
\[
(x,\vf^{\ga},A_{i_{1_{1}}}^{1\ga},\dots,A_{i_{(k-2)_{1}}\dots i_{(k-2)_{k-2}}}^{(k-2)\ga};\vf_{,j_{0}}^{\ga},A_{i_{1_{1}},j_{1}}^{1\ga},\dots,A_{i_{(k-2)_{1}}\dots i_{(k-2)_{k-2}},j_{k-2}}^{(k-2)\ga})
\]
 and an element of $V^{0}J^{1}U\isom L(T\spc,J^{k-2}\vbts)$, for
which $\vf^{\ga}=0$, $A_{i_{1_{1}}}^{1\ga}=0$, \ldots{}, $A_{i_{(k-2)_{1}}\dots i_{(k-2)_{k-2}}}^{(k-2)\ga}=0$,
is of the form 
\[
(x,\vf_{,j_{0}}^{\ga},A_{i_{1_{1}},j_{1}}^{1\ga},\dots,A_{i_{(k-2)_{1}}\dots i_{(k-2)_{k-2}},j_{k-2}}^{(k-2)\ga}).
\]
It follows that, $\incl_{V}^{0*}(Y)$ is of the form
\begin{equation}
(x,Y_{1\dots\wh{\imath}_{r_{0}}\dots n\ga}^{0j_{0}},Y_{1\dots\wh{\imath}_{r_{1}}\dots n\ga}^{1i_{1_{1}}j_{1}},\dots,Y_{1\dots\wh{\imath}_{r_{k-2}}\dots n\ga}^{(k-2)i_{(k-2)_{1}}\dots i_{(k-2)_{k-2}}j_{k-2}}).
\end{equation}
In the last equation, the arrays $Y^{p}$ are symmetric with respect
to the indices $i_{p_{1}},\dots,i_{p_{p}}$.

\end{rem}

\subsection{Invariant components of non-holonomic hyper-stresses\label{subsec:Signifcant-components-NHHS}}

The inclusion mappings of the various vertical subbundles of $J^{1}(J^{1}\avb)$
make it possible to consider the induced projections, yielding invariant
components of non-holonomic hyper-stresses. Thus, we have the projections
\begin{equation}
\begin{split}\incl_{123}^{*}:L(J^{1}(J^{1}\avb),\ext^{n}T^{*}\spc) & \tto L(V_{123}J^{1}(J^{1}\avb),\ext^{n}T^{*}\spc),\\
\incl_{13}^{*}:L(J^{1}(J^{1}\avb),\ext^{n}T^{*}\spc) & \tto L(V_{13}J^{1}(J^{1}\avb),\ext^{n}T^{*}\spc),\\
\incl_{23}^{*}:L(J^{1}(J^{1}\avb),\ext^{n}T^{*}\spc) & \tto L(V_{23}J^{1}(J^{1}\avb),\ext^{n}T^{*}\spc),
\end{split}
\end{equation}
all given by compositions with the corresponding inclusions. Locally,
these projections are given by
\begin{equation}
\begin{split}(x^{i},X_{1\dots n\ga_{0}}^{0},X_{1\dots n\ga_{1}}^{1i_{1}},X_{1\dots n\ga_{2}}^{2i_{2}},X_{1\dots n\ga_{3}}^{3i_{3}i_{4}}) & \longmapsto(x^{i},X_{1\dots n\ga_{1}}^{1i_{1}},X_{1\dots n\ga_{2}}^{2i_{2}},X_{1\dots n\ga_{3}}^{3i_{3}i_{4}}),\\
(x^{i},X_{1\dots n\ga_{0}}^{0},X_{1\dots n\ga_{1}}^{1i_{1}},X_{1\dots n\ga_{2}}^{2i_{2}},X_{1\dots n\ga_{3}}^{3i_{3}i_{4}}) & \longmapsto(x^{i},X_{1\dots n\ga_{1}}^{1i_{1}},,X_{1\dots n\ga_{3}}^{3i_{3}i_{4}}),\\
(x^{i},X_{1\dots n\ga_{0}}^{0},X_{1\dots n\ga_{1}}^{1i_{1}},X_{1\dots n\ga_{2}}^{2i_{2}},X_{1\dots n\ga_{3}}^{3i_{3}i_{4}}) & \longmapsto(x^{i},X_{1\dots n\ga_{2}}^{2i_{2}},X_{1\dots n\ga_{3}}^{3i_{3}i_{4}}),
\end{split}
\end{equation}
respectively.

Of particular interest is the projection
\begin{equation}
\incl_{3}^{*}:L(J^{1}(J^{1}\avb),\ext^{n}T^{*}\spc)\tto L(V_{3}J^{1}(J^{1}\avb),\ext^{n}T^{*}\spc)\isom L(L^{2}(T\spc,\avb),\ext^{n}T^{*}\spc),
\end{equation}
represented locally by 
\begin{equation}
(x^{i},X_{1\dots n\ga_{0}}^{0},X_{1\dots n\ga_{1}}^{1i_{1}},X_{1\dots n\ga_{2}}^{2i_{2}},X_{1\dots n\ga_{3}}^{3i_{3}i_{4}})\longmapsto(x^{i},X_{1\dots n\ga_{3}}^{3i_{3}i_{4}}).
\end{equation}
We will refer to elements in the image of $\incl_{3}^{*}$ as the
significant components of the corresponding non-holonomic stresses.
Evidently, we have a natural inclusion of significant components of
second order hyper-stresses (as in Section \ref{subsec:Significant-components-HS})
in the collection of significant components of non-holonomic hyper-stresses
and a natural projection given by symmetrization. Specifically, the
significant components of a non-holonomic stress are of the form
\begin{equation}
\sum_{\ga,i,j}X_{1\dots n\ga}^{3ij}\partial_{i}\otimes\partial_{j}\otimes g^{\ga}\otimes(\lisupc{\dee x}{\wedge}n)\label{eq:rep-signif-comp-nhhs}
\end{equation}
and for second order stresses, $X_{1\dots n\ga}^{3ij}$ are symmetric
in the $i,j$ indices.

\subsection{Contraction operations on significant components\label{subsec:Contraction-operations-on}}

In analogy with the contraction mapping defined in Section \ref{sec:Simple-Stresses},
as a particular case of Remark \ref{rem:Contraction}, one can define
a contraction mapping 
\begin{equation}
\begin{split}\spec C^{1}:\L{L^{2}(T\spc,\avb),\ext^{n}T^{*}\spc} & \tto L(L(T\spc,\avb),\ext^{n-1}T^{*}\spc),\\
 & \,\,\,\,\,\,\,\,\isom T\spc\otimes\avb^{*}\otimes\ext^{n}T^{*}\spc.
\end{split}
\end{equation}
To this end, one can simply use the identification $L^{2}(T\spc,\avb)\isom L(T\spc,L(T\spc,\avb))$
and use the definition in (\ref{eq:contraction-map}) and (\ref{eq:contraction-defined})
for $\vbts=L(T\spc,\avb)$. Consider the composition $\spec C^{1}\comp\incl_{3}^{*}$.
For $X\in L(J^{1}(J^{1}\avb),\ext^{n}T^{*}\spc)$ represented as in
(\ref{eq:rep-nhhs}), $\spec C^{1}\comp\incl_{3}^{*}(X)$ is represented
by
\begin{multline}
\sum_{\ga,i,j}X_{1\dots n\ga}^{3ij}\partial_{j}\otimes g^{\ga}\otimes\left(\partial_{i}\contr(\lisupc{\dee x}{\wedge}n)\right)\\
=\sum_{\ga,i,j}(-1)^{i-1}X_{1\dots n\ga}^{3ij}\partial_{j}\otimes g^{\ga}\otimes(\lisuppwout{\dee x}1{\wedge}{\cdots}ni).\label{eq:rep-first-reduction}
\end{multline}

Let $X:\spc\to\L{J^{1}(J^{1}\avb),\ext^{n}T^{*}\spc}$ be a non-holonomic
stress field and let $\var$ be an $(n-1)$-dimensional submanifold
of $\spc$. Then, the inclusion $\incl_{\var}:\var\to\spc$ induces
a restriction $\incl_{\var}^{*}$ of $(n-1)$-forms onto $\var$ and
so we have a field 
\begin{equation}
\sst=\incl_{\var}^{*}\comp\spec C^{1}\comp\incl_{3}^{*}\comp X\in L(L(T\spc,\avb),\ext^{n-1}T^{*}\var),
\end{equation}
the hyper-surface stress induced by the hyper-stress. It is noted
that the surface stress is not ``tangent'' to the surface as it acts
on elements of $L(T\spc,\avb)$ and not elements of $L(T\var,\avb)$.
In particular, one cannot perform a meaningful second contraction
on $\sst$ because of this reason.

\subsection{Second contraction\label{subsec:Second-contraction}}

Alluding to the last remark, one can perform a second contraction
on $\spec C^{1}\comp\incl_{3}^{*}(X)$ to obtain an $(n-2)$-vector
valued form $\spec C\comp\spec C^{1}\comp\incl_{3}^{*}(X)$ in $\spc$.
Using the representation in (\ref{eq:rep-first-reduction}), $\spec C\comp\spec C^{1}\comp\incl_{3}^{*}(X)$
is represented by
\begin{equation}
\sum_{\ga,i,j}(-1)^{i-1}X_{1\dots n\ga}^{3ij}g^{\ga}\otimes\left(\partial_{j}\contr(\lisuppwout{\dee x}1{\wedge}{\cdots}ni)\right)
\end{equation}
This may seem promising because such a form, valued in $\avb^{*}$,
would induce edge forces, forces concentrated on submanifolds of dimension
$(n-2)$. However, as we show below, this form vanishes identically
for symmetric (compatible) significant components. Using
\begin{multline}
\parder[x^{j}]{}\contr(\lisuppwout{\dee x}1{\wedge}{\cdots}ni)=\\
\begin{cases}
(-1)^{j-1}\dee x^{1}\wedge\cdots\wedge\wh{\dee x^{j}}\wedge\cdots\wedge\wh{\dee x^{i}}\wedge\cdots\wedge\dee x^{n}, & \text{for }j<i,\\
-(-1)^{j-1}\dee x^{1}\wedge\cdots\wedge\wh{\dee x^{i}}\wedge\cdots\wedge\wh{\dee x^{j}}\wedge\cdots\wedge\dee x^{n}, & \text{for }j>i,
\end{cases}
\end{multline}
it follows that the representation of $\spec C\comp\spec C^{1}\comp\incl_{3}^{*}(X)$
may be rewritten as

\begin{equation}
\sum_{\ga,i>j}(-1)^{i+j}(X_{1\dots n\ga}^{3ij}-X_{1\dots n\ga}^{3ji})g^{\ga}\otimes(\dee x^{1}\wedge\cdots\wedge\wh{\dee x^{j}}\wedge\cdots\wedge\wh{\dee x^{i}}\wedge\cdots\wedge\dee x^{n}).
\end{equation}
Since we expect that for the significant components of second order
non-holonomic hyper-stresses, $X_{1\dots n\ga}^{3ij}=X_{1\dots n\ga}^{3ji}$,
the form vanishes identically. This is of course expected as we contract
twice a completely anti-symmetric tensor with a symmetric 2-tensor.

\subsection{The tangent projection of jets and vertical jets for a given submanifold\label{subsec:Vertical-jets}}

Consider $(J^{1}\avb)\resto{\var}=\incl_{\var}^{*}J^{1}\avb$ for
an $(n-1)$-dimensional submanifold $\var$ of $\spc$ and the jet
bundle $J^{1}(\avb\resto{\var})=J^{1}(\incl_{\var}^{*}\avb)$. One
has a natural projection 
\begin{equation}
\proj:(J^{1}\avb)\resto{\var}\tto J^{1}(\avb\resto{\var}),\quad\text{given by}\quad A=j^{1}\avf(y)\longmapsto j^{1}(\avf\resto{\var}(y)),
\end{equation}
for any section $\avf$ of $\avb$ that represents $A$. Locally,
$\proj$ is represented by
\begin{equation}
(y^{a},\avf^{\ga},A_{i}^{\gb})\lmt(y^{a},u^{\ga},B_{b}^{\gb}),\quad i=1,\dots,n,\quad b=1,\dots,n-1,
\end{equation}
where $B_{b}^{\gb}=A_{b}^{\gb}$ in an adapted coordinate system.

Evidently, $\proj$ is linear and we may consider
\begin{equation}
\proj^{*}:\L{J^{1}(\avb\resto{\var}),\ext^{n-1}T^{*}\var}\tto\L{(J^{1}\avb)\resto{\var},\ext^{n-1}T^{*}\var},\quad Z\lmt Z\comp\proj.
\end{equation}
Locally, if $Z\in\L{J^{1}(\avb\resto{\var}),\ext^{n-1}T^{*}\var}$
is represented by $(y^{a},Z_{\ga}^{0},Z_{\gb}^{1b})$, then, $\proj^{*}(Z)$
is represented by $(y^{a},Z_{\ga}^{0},Z_{\gb}^{1i})$ with $Z_{\gb}^{1n}=0$.

We will say that $A\in(J^{1}\avb)\resto{\var}=\incl_{\var}^{*}J^{1}\avb$
is \emph{vertical} if $\proj\,A=0$, so that $A$ is represented by
a section which vanishes at $\pi^{1}A$ together with its tangential
derivatives. The vertical jets form the \emph{vertical subbundle}
$V_{\var}((J^{1}\avb)\resto{\var})=\kernel\proj$. Let $y^{a}$, $a=1,\dots,n-1$,
be coordinates in a local chart on $\var$ so that $y^{1},\dots,y^{n-1},x^{n}$,
is a chart in $\spc$. If $(y^{a},\avf^{\ga},A_{b}^{\gb},A_{n}^{\gb})$
represent an element $A\in(J^{1}\avb)\resto{\var}$, then, $A$ belongs
to the vertical subbundle if and only if $\avf^{\ga}=0$ and $A_{b}^{\gb}=0$,
$b=1,\dots,n-1$. Evidently, the fiber $V_{\var}((J^{1}\avb)\resto{\var})\resto y$
is isomorphic to $L(\reals,\avb_{y})\isom\avb_{y}$. 

\begin{rem}
\textbf{The normal bundle. }While the isomorphism of $V_{\var}((J^{1}\avb)\resto{\var})\resto y$
with $L(\reals,\avb_{y})\isom\avb_{y}$ depends on the chart, one
may construct a natural isomorphism of $V_{\var}((J^{1}\avb)\resto{\var})$
with $\L{\Xi,\avb\resto{\var}}$. Here, $\Xi$ is the subbundle of
$T^{*}\spc\resto{\var}$ containing the annihilators of $T\var$,
$\Xi_{y}=T_{y}\var^{\perp}$, \ie,  the one-dimensional subspace
of forms $\xi$, with $\xi(v)=0$ for all $v\in T_{y}\var$. Let $\rho:T^{*}\spc\to T^{*}\var$
be the natural restriction. Then, $\vph\in\kernel\rho$ if and only
if $\rho(\vph)(v)=\vph(v)=0$ for all $v\in T_{y}\var$. Hence, we
have the identification $\Xi\isom\kernel\rho$.

Let $\incl_{0\Xi}:\Xi\to T^{*}\spc$ be the natural inclusion. Then,
we have the surjection $\incl_{0\Xi}^{*}:T\spc\to\Xi^{*}$ and $\incl_{0\Xi}^{*}(v)=\incl_{0\Xi}^{*}(v')$
if and only if $\vph(v-v')=0$ for any $\vph\in\Xi$, \ie, $v-v'\in T\var$.
Thus, each element of $\Xi^{*}$ determines a unique element of $T\spc/T\var$,
and one makes the identification $\Xi^{*}\isom T\spc/T\var$. In fact,
$\incl_{0\Xi}^{*}:T\spc\to\Xi^{*}\isom T\spc/T\var$ is simply the
natural projection on the quotient.
\end{rem}

 To construct an isomorphism 
\begin{equation}
\incl_{\Xi}:V_{\var}((J^{1}\avb)\resto{\var})\tto\L{\Xi,\avb\resto{\var}},\label{eq:IsomLayers}
\end{equation}
we first note that $V_{\var}((J^{1}\avb)\resto{\var})$ is a subbundle
of $(VJ^{1}\avb)\resto{\var}\isom\L{T\spc,\avb}\resto{\var}$. In
fact, while $\L{T\spc,\avb}\resto{\var}=(T^{*}\spc\otimes\avb)\resto{\var}$,
\begin{equation}
V_{\var}((J^{1}\avb))_{y}=T_{y}\var^{\perp}\otimes\avb_{y}=\L{\Xi_{y},\avb_{y}}
\end{equation}
as vertical jets are exactly those elements for which the actions
on tangent vectors vanish. 

One may consider therefore the inclusion
\begin{equation}
\incl_{V_{\var}}:V_{\var}((J^{1}\avb)\resto{\var})\tto J^{1}\avb\resto{\var}
\end{equation}
and the \emph{vertical projection} of surface stresses
\begin{equation}
\incl_{V_{\var}}^{*}:L((J^{1}\avb)\resto{\var},\ext^{n-1}T^{*}\var)\tto L(V_{\var}((J^{1}\avb)\resto{\var}),\ext^{n-1}T^{*}\var),\qquad Z\longmapsto Z\comp\incl_{V_{\var}}.\label{eq:vertProj}
\end{equation}
If $Z\in L((J^{1}\avb)\resto{\var},\ext^{n-1}T^{*}\var)$ is represented
by $(y^{b},Z_{1\dots n-1\ga}^{0},Z_{1\dots n-1\gb}^{1b},Z_{1\dots n-1\gb}^{1n})$
so that 
\begin{equation}
Z(A)=\left[\sum_{\ga}\left(Z_{1\dots n-1\ga}^{0}u^{\ga}+Z_{1\dots n-1\ga}^{1n}A_{n}^{\ga}\right)+\sum_{\ga,b}Z_{1\dots n-1\ga}^{1b}A_{b}^{\ga}\right]\lisuppc{\dee y}1{\wedge}{n-1},\label{eq:repSurStress}
\end{equation}
then, $\incl_{V_{\var}}^{*}(Z)$ is of the form $(y^{b},Z_{1\dots n-1\gb}^{1n})$,
which in view of (\ref{eq:IsomLayers}) is a representation of an
element in $\L{\L{\Xi,\avb\resto{\var}},\ext^{n-1}T^{*}\var}\isom\Xi\otimes\avb^{*}\resto{\var}\otimes\ext^{n-1}T^{*}\var$.

It is observed that one could define the vertical projection somewhat
differently (though we keep the same notation) as 
\begin{equation}
\begin{split}\incl_{V_{\var}}^{*}:L((J^{1}\avb)\resto{\var},(\ext^{n-1}T^{*}\spc)\resto{\var}) & \tto L(V_{\var}((J^{1}\avb)\resto{\var}),(\ext^{n-1}T^{*}\spc)\resto{\var}),\\
 & \qquad\isom\Xi\otimes\avb^{*}\resto{\var}\otimes(\ext^{n-1}T^{*}\spc)\resto{\var},
\end{split}
\label{eq:transverseProjectnOfStresses}
\end{equation}
which could then be composed on the left with the restriction of forms.
These observations can summarized by the sequences

\begin{equation}\begin{CD}
0\tto  V_{\var}((J^{1}\avb)\resto{\var}) @>{\incl_{V_{\var}}}>>  (J^{1}\avb)\resto{\var} @>\proj>>J^{1}(\avb\resto{\var}) \tto 0 \\
0\longleftarrow V_{\var}((J^{1}\avb)\resto{\var})^* @<{\incl^*_{V_{\var}}}<<  (J^{1}\avb)^*\resto{\var} @<\proj^*<<J^{1}(\avb\resto{\var})^* \longleftarrow 0,
\end{CD}\label{eq:CDa}\end{equation}where in the second sequence we wrote to dual bundles, rather than
the linear maps into $\ext^{n-1}T^{*}\var$, for short.

We conclude that the vertical projection $\incl_{V_{\var}}^{*}(Y)$
of a hyper-surface stress $Y=p_{\st}(X)$ has an invariant meaning.
It is somewhat reminiscent of the bending moment in shell theory.

The restriction of the diagram in (\ref{eq:CDa}) to the vertical
subbundle is of the form 

\begin{equation}\begin{CD}
0\tto  V_{\var}(V(J^{1}\avb)\resto{\var}) @>{\incl_{V_{\var}}}>>  (VJ^{1}\avb)\resto{\var} @>\proj>>VJ^{1}(\avb\resto{\var}) \tto 0 \\
0\longleftarrow V_{\var}(V(J^{1}\avb)\resto{\var})^* @<{\incl^*_{V_{\var}}}<<  (VJ^{1}\avb)^*\resto{\var} @<\proj^*<<VJ^{1}(\avb\resto{\var})^* \longleftarrow 0,
\end{CD}\label{eq:CDaa}\end{equation}or,

\begin{equation}\begin{CD}
0\tto  \Xi\otimes\avb\resto\var @>{\incl_{V_{\var}}}>>  (T^*\spc\otimes\avb)\resto{\var} @>\proj>>T^*\var\otimes\avb\resto{\var} \tto 0 \\
0\longleftarrow \Xi^*\otimes{\avb^*\resto\var} @<{\incl^*_{V_{\var}}}<<  (T\spc\otimes\avb^*)\resto{\var} @<\proj^*<<T\var\otimes\avb\resto{\var}^* \longleftarrow 0,
\end{CD}\label{eq:CDaa}\end{equation}where we have not indicated the restrictions for the various mappings.

\subsection{Tangent surface stresses\label{subsec:Tangent-surface-stresses}}

We say that a surface stress $Z\in L((J^{1}\avb)\resto{\var},\ext^{n-1}T^{*}\var)$
is tangent to the submanifold $\var$ if its vertical component vanishes.
That is, $\incl_{V_{\var}}^{*}(Z)=0$, or alternatively, $Z_{1\dots n-1\gb}^{1n}=0$
for all $\gb$. From Equation (\ref{eq:CDa}), $Z$ is tangent to
$\var$ when it is in the image of $\proj^{*}$. It is emphasized
that there is no natural projection of surface stresses onto tangent
surface stresses.

Let $Z\in L((J^{1}\avb)\resto{\var},\ext^{n-1}T^{*}\var)$ be a surface
stress tangent to $\var$ at some point $y\in\var$. Then, for a section
$\vf$ of $\avb$, and using the notation above, 
\begin{equation}
Z(j^{1}\vf(y))=\left[\sum_{\ga}Z_{1\dots n-1\ga}^{0}u^{\ga}+\sum_{\ga,b}Z_{1\dots n-1\ga}^{1b}u_{,b}^{\ga}\right]\lisuppc{\dee y}1{\wedge}{n-1}.
\end{equation}
It follows that $Z$ determines linearly a unique element of $L(J^{1}(\avb\resto{\var}),\ext^{n-1}T^{*}\var)$
so that there is a natural isomorphism of the subbundle of tangent
surface stresses with $L(J^{1}(\avb\resto{\var}),\ext^{n-1}T^{*}\var)$.

Consider the term
\begin{equation}
I_{1}=\int_{\bdry\body}Y(A)\label{eq:BoudaryIntegralToComput}
\end{equation}
appearing in Equation (\ref{eq:firstIntegrationByParts}). Here $Y$
is a section of $L(J^{1}\avb,\ext^{n-1}T^{*}\spc)$ and we consider
the case where the section $A$ of $J^{1}U$ is compatible, that is,
there is a section $\avf$ of $\avb$ with $A=j^{1}\avf$. Let $Z$
be the restriction of $Y$ to $\bdry\body$ so that 
\begin{equation}
I_{1}=\int_{\bdry\body}Z(j^{1}\avf).
\end{equation}
Assuming that $\bdry\body$ is piecewise smooth, let $\var\subset\bdry\body$
be a smooth $(n-1)$-dimensional submanifold. If $Z$ is tangent to
$\var$, one may use (\ref{eq:DefineDivergence-1}) for $\vbts=\avb\resto{\var}$
and obtain
\begin{equation}
\begin{split}\int_{\var}Z(j^{1}\avf) & =\int_{\var}\dee(p_{\st}(Z)(\avf))-\int_{\var}\diver Z(\avf),\\
 & =\int_{\bdry\var}p_{\st}(Z)(\avf)-\int_{\var}\diver Z(\avf),
\end{split}
\end{equation}
where $p_{\st}(Z)$ is a section of $L(\avb\resto{\var},\ext^{n-2}T^{*}\var)$,
\ie, a vector valued $(n-2)$-form on $\var$. Thus, the restriction
of $p_{\st}(Z)$ to $\bdry\var$ induces an edge force on $\bdry\var$.

\section{Additional Geometric Structure\label{sec:Additional-Geometric-Structure}}

Our objective in this section is to introduce sufficient geometric
structure so that a hyper-surface stress may be decomposed into tangent
and transverse components relative to $\var$. Such a decomposition
will make it possible to determine unique edge forces which are induced
by the non-holonomic hyper-stress. 

In analogy with (\ref{eq:contraction-map}) one would attempt to
use contraction on $T\spc\resto{\var}\otimes\ext^{n-1}T^{*}\var$.
However, this cannot be done because there is no natural extension
of forms from $\ext^{n-1}T^{*}\var$ to $\ext^{n-1}T^{*}\spc$. 

Assuming that $\var$ is orientable, let $\nor$ be a nowhere vanishing
vector field in $T\spc\resto{\var}$ which is transversal to $T\var$
and let $N$ be the induced transverse bundle, that is, $N_{y}=\{a\nor(y)\mid a\in\reals\}$.
Evidently, if $\spc$ is a Riemannian manifold, the metric induces
such a transversal, normal, field. Thus, we have a decomposition
\begin{equation}
T\spc\resto{\var}\isom T\var\oplus N.\label{eq:TangentSpaceDecomp}
\end{equation}
Let $\vph_{\nor}$ be the 1-form on $T\spc$ that annihilates $T\var$
such that $\vph_{\nor}(\nor)=1$ and let $\proj_{1}$ and $\proj_{2}$
be the projections giving the tangent and transverse components of
the decomposition. Then, 
\begin{equation}
\proj_{1}=\idnt-\vph_{\nor}\otimes\nor,\qquad\proj_{2}=\vph_{\nor}\otimes\nor
\end{equation}
and we will use the notation $v_{\var}=\proj_{1}(v)$, $v_{N}=\proj_{2}(v)$.
Using adapted coordinates, $(v_{\var})^{a}=\sum_{i}(\gd_{i}^{a}-\vph_{i}n^{a})v^{i}$,
$(v_{N})^{i}=\sum_{j}\vph_{j}v^{j}n^{i}$, where $\vph_{j}$ are the
components of $\vph_{\nor}$. It is observed that $\proj_{2}$ induces
an isomorphism 
\begin{equation}
\Xi^{*}\isom T\spc/T\var\tto N,\qquad[v]\mapsto\proj_{2}(v)=\vph_{\nor}(v)\nor.\label{eq:Isom_Layers_Normals}
\end{equation}

Next, we introduce some notation. Evidently, $(T\var\oplus N)^{*}\isom T^{*}\var\oplus N^{*}$.
In fact, if $\incl_{1}:T\var\to T\spc$ and $\incl_{2}:N\to T\spc$
are the natural inclusions, then, the projections $\incl_{1}^{*}:T^{*}\spc\to T^{*}\var$
and $\incl_{2}^{*}:T^{*}\spc\to N^{*}$ are the restrictions of forms
and $\proj_{1}^{*}=\idnt^{*}-\nor\otimes\vph_{\nor}$, $\proj_{2}^{*}=\nor\otimes\vph_{\nor}$.
In particular, for every $\psi\in T^{*}\spc$, it is easy to verify
that $\psi=\psi_{\var}+\psi_{N}$, where $\psi_{\var}=\psi-\psi(\nor)\vph_{\nor}=(\idnt^{*}-\nor\otimes\vph_{\nor})(\psi)$
and $\psi_{N}=\psi(\nor)\vph_{\nor}=(\nor\otimes\vph_{\nor})(\psi)$.
In addition $\psi_{\var}(v_{N})=0$, $\psi_{N}(v_{\var})=0$ so that
$\psi(v)=\psi_{\var}(v_{\var})+\psi_{N}(v_{N})$. In adapted components,
$\psi_{\var}=\sum_{a}(\psi_{\var})_{a}\dee y^{a}$, where $(\psi_{\var})_{a}=\psi_{a}-\sum_{j}\psi_{j}n^{j}\vph_{a}$.

It follows that 
\begin{equation}
\L{T\spc,\avb}\resto{\var}\isom T^{*}\spc\resto{\var}\otimes\avb\resto{\var}\isom T^{*}\var\otimes\avb\resto{\var}\oplus N^{*}\otimes\avb\resto{\var}\isom\L{T\var,\avb\resto{\var}}\oplus\L{N,\avb\resto{\var}},
\end{equation}
and 
\begin{equation}
\begin{split}L(\L{T\spc,\avb}\resto{\var},\ext^{n-1}T^{*}\var) & \isom L(T\var,\avb\resto{\var}),\ext^{n-1}T^{*}\var)\oplus L(\L{N,\avb\resto{\var}},\ext^{n-1}T^{*}\var),\\
 & \isom T\var\otimes\avb^{*}\resto{\var}\otimes\ext^{n-1}T^{*}\var\oplus N\otimes\avb^{*}\resto{\var}\otimes\ext^{n-1}T^{*}\var.
\end{split}
\label{eq:DecomposeSurfaceStress}
\end{equation}
We will keep the notation $\proj_{1}$ and $\proj_{2}$ for the two
projections of the product in (\ref{eq:DecomposeSurfaceStress}).
The first component, $\overline{Y}:=\proj_{1}(\incl_{V}^{*}(Y))=(\incl_{V}^{*}(Y))_{\var}\in T\var\otimes\avb^{*}\resto{\var}\otimes\ext^{n-1}T^{*}\var$,
is represented locally by 
\begin{equation}
\sum_{a,\ga}\overline{Y}_{1\dots n-1\ga}^{1a}\partial_{a}\otimes g^{\ga}\otimes(\lisuppc{\dee y}1{\wedge}{n-1}),
\end{equation}
with 
\begin{equation}
\overline{Y}_{1\dots n-1\ga}^{1a}=\sum_{i}Y_{1\dots n-1\ga}^{1i}(\gd_{i}^{a}-\vph_{i}n^{a}),\label{eq:Comp_Ybar}
\end{equation}
 and $\proj_{2}(Y)=\incl_{V_{\var}}^{*}(\incl_{V}^{*}(Y))_{N}\in N\otimes\avb^{*}\resto{\var}\otimes\ext^{n-1}T^{*}\var$
is represented in the form
\begin{multline}
\sum_{i,j,\ga}Y_{1\dots n-1\ga}^{1i}\vph_{i}n^{j}\bdry_{j}\otimes g^{\ga}\otimes(\lisuppc{\dee y}1{\wedge}{n-1})\\
=\sum_{i,j,\ga}Y_{1\dots n-1\ga}^{1i}\vph_{i}\nor\otimes g^{\ga}\otimes(\lisuppc{\dee y}1{\wedge}{n-1}).
\end{multline}
The situation is illustrated in the following diagram.

\begin{equation}
\begin{diagram}[p=3pt,w=2em,h=3.5em]
0 &\pile{\rTo^{}\\ \\ \lTo_{}}  &{N^*\otimes\avb\resto\var} 
&  \pile{\rTo^{{\;\proj_2^*\comp\incl_{V_{\var}}}\;\;}\\ \\ \lTo_{\rho}} 
       &{(T^*\spc\otimes\avb)\resto{\var}}
&  \pile{\rTo^{{\;\;\;\proj}\;\;\;}\\ \\ \lTo_{\proj_1^*}} 
       &{T^*\var\otimes\avb\resto{\var} }
&  \pile{\rTo^{}\\ \\ \lTo_{}} 
       &0,
\\
0 &\pile{\lTo^{}\\ \\ \rTo_{}}  &N\otimes\avb^*\resto\var 
&  \pile{\lTo^{{\;\;\;\proj_2\comp\incl_{V_{\var}}^*}\;\;}\\ \\ \rTo_{\incl_N}} 
       &{(T\spc\otimes\avb^*)\resto{\var}}
&  \pile{\lTo^{{\;\;\;\proj^*}\;\;\;}\\ \\ \rTo_{\proj_1}} 
       &{T\var\otimes\avb\resto{\var}^*}
&  \pile{\rTo^{}\\ \\ \lTo_{}} 
       &0,
\end{diagram}
\end{equation}where $\incl_{N}$ and $\rho$ are the natural inclusion of $N$ and
its dual, respectively; $\proj_{2}$ is the isomorphism of (\ref{eq:Isom_Layers_Normals}).

\begin{rem}
One might try to decompose a jet into two components, one tangent
to $\var$ and the second normal to $\var$. The two projections are
well defined. However, the projections do not endow $(J^{1}\avb)\resto{\var}$
with a structure of a direct sum. This is because none of the spaces
$J^{1}(\avb\resto{\var})$ and $J_{\nor}^{1}(\avb\resto{\var})$ is
a subspace of $(J^{1}\avb)\resto{\var}$, nor can it be made naturally
isomorphic to a sub-bundle. For example, if we were able to identify
$J^{1}(\avb\resto{\var})$ with a sub-bundle of $(J^{1}\avb)\resto{\var}$,
we would be able to extend jets and we could restrict stresses to
$\var$.
\end{rem}

It is now possible to perform contraction on the first component $\incl_{V}^{*}(Y)_{\var}\in T\var\otimes\avb^{*}\resto{\var}\otimes\ext^{n-1}T^{*}\var$
and obtain 
\begin{equation}
\ssp_{\var}(Y)=\spec C(\incl_{V}^{*}(Y)_{\var})\in\avb^{*}\resto{\var}\otimes\ext^{n-2}T^{*}\var\isom\L{U\resto{\var},\ext^{n-2}T^{*}\var}
\end{equation}
which is represented in analogy with (\ref{eq:contraction-defined})
as
\begin{equation}
\sum_{a,\ga}(-1)^{a-1}\overline{Y}_{1\dots n-1\ga}^{1a}g^{\ga}\otimes(\lisuppwout{\dee y}1{\wedge}{\cdots}{n-1}a).
\end{equation}
Substituting (\ref{eq:Comp_Ybar}) and (\ref{eq:Rep-PsigmaX-1}),
the representation of $\ssp_{\var}(p_{\st}(X))$ is given in an adapted
coordinate system by
\begin{equation}
\sum_{i,a,\ga}(-1)^{a-1}(-1)^{n-1}X_{1\dots n\ga}^{3in}(\gd_{i}^{a}-\vph_{i}n^{a})g^{\ga}\otimes(\lisuppwout{\dee y}1{\wedge}{\cdots}{n-1}a).
\end{equation}
The field $\ssp_{\var}(Y)$ represents tangent surface traction
stress, as expected. 

The construction leading to the determination of the tangent surface
traction (hyper-) stress is summarized in the following diagram..

\begin{equation}
\begin{xy}
(30,40)*+{{\L{(J^1\avb)\resto{\var},\ext^{n-1}T^*\var}}}="o";
(0,20)*+{{T\spc\resto{\var}\otimes{\avb^*}\resto{\var}\otimes\ext^{n-1}T^*\var}}="a";
(60,0)*+{{N\otimes\avb^*\resto{\var}\ext^{n-1}T^*\var}}="b";%
(60,20)*+{{\Xi^*\otimes\avb^*\resto{\var}\ext^{n-1}T^*\var}}="x";%
(0,0)*+{{T\var\otimes\avb^*\resto{\var}\otimes\ext^{n-1}T^*\var}}="c"; 
(0,-20)*+{{\avb^*\resto{\var}\otimes\ext^{n-2}T^*\var}}="e";%
{\ar "x";"b"}?*!/_1mm/{{{\qquad\proj_2}}};
{\ar "o";"a"};?*!/^4mm/{\incl_V^*};
{\ar "o";"x"};?*!/_4mm/{\incl_{V_\var}^*};
{\ar "a";"c"};?*!/^2mm/{\proj_1\quad};
{\ar@{.>} "a";"x"};?*!/^4mm/{};
{\ar "c";"e"};?*!/^4mm/{\spec{C}};
{\ar@/^2.5pc/ "o";"e"};?*!/_3mm/{p_\tau};
{\ar@/_2pc/ "o";"b"};?*!/^3mm/{p_\nu};
\end{xy}
\end{equation}
\smallskip 

The generalized surface divergence of $Y$, a section of $\L{J^{1}\avb\resto{\var},\ext^{n-1}T^{*}\var}$
is defined now by 

\begin{equation}
\sdiv_{\var}Y(j^{1}\avf)=\dee(\ssp_{\var}(Y)(u))-Y(j^{1}\avf).\label{eq:SurfDiv}
\end{equation}
With the representation of $\dee(\ssp_{\var}(Y)(\avf))$ by 
\begin{equation}
\left(\sum_{\ga,a}\overline{Y}_{1\dots n-1\ga,a}^{1a}\avf^{\ga}+\sum_{\ga,a}\overline{Y}_{1\dots n-1\ga}^{1a}\avf_{,a}^{\ga}\right)\lisuppc{\dee y}1{\wedge}{n-1},
\end{equation}
and the representation of $Y(j^{1}\avf)$ by 
\begin{equation}
\left(\sum_{\ga}Y_{1\dots n-1\ga}^{0}\avf^{\ga}+\sum_{\ga,a}\overline{Y}_{1\dots n-1\ga}^{1a}\avf_{,a}^{\ga}+\sum_{\ga,i,j}Y_{1\dots n-1\ga}^{1i}\vph_{i}\avf_{,j}^{\ga}n^{j}\right)\lisuppc{\dee y}1{\wedge}{n-1},
\end{equation}
$\sdiv_{\var}Y(j^{1}u)$ is represented by 
\begin{equation}
\left(\sum_{\ga,a}\overline{Y}_{1\dots n-1\ga,a}^{1a}\avf^{\ga}-\sum_{\ga}Y_{1\dots n-1\ga}^{0}\avf^{\ga}-\sum_{\ga,i,j}Y_{1\dots n-1\ga}^{1i}\vph_{i}\avf_{,j}^{\ga}n^{j}\right)\lisuppc{\dee y}1{\wedge}{n-1},
\end{equation}
which evidently depends on the derivative of the vector field in the
direction of the transverse vector $\nor$. This is of course is the
reason why the divergence of the hyper-surface stress acts on the
jet of generalized velocity $\avf$ rather than the values of the
field itself as in the case of simple stresses.

\begin{rem}
One may attempt to view the term $\sum_{\ga,a,j}Y_{1\dots n-1\ga}^{1i}\vph_{i}\avf_{,j}^{\ga}n^{j}$
as the action of $\incl_{V_{\var}}^{*}(Y)$. However, as opposed to
(\ref{eq:transverseProjectnOfStresses}), $\avf_{,j}^{\ga}$ cannot
be viewed as components of an element in $\Xi\otimes\avb^{*}\resto{\var}$.
It is only the additional structure induced by the vector field $\nor$
that induces the annihilator $\vph_{i}u_{,j}^{\ga}n^{j}$. (It is
true, of course, that any other annihilator of $T\var$ may be expressed
as a product of $\vph_{i}\avf_{,j}^{\ga}n^{j}$ with a scalar valued
function.)
\end{rem}

\begin{rem}
Evidently, the derivatives 
\[
\overline{Y}_{1\dots n-1\ga,a}^{1a}=\sum_{i}\left[Y_{1\dots n-1\ga}^{1i}(\gd_{i}^{a}-\vph_{i}n^{a})\right]_{,a}
\]
 include the derivatives of the transverse vector field $\nor$ and
the associated form $\vph$ which indicate the ``curvature'' associated
with the field $\nor$.
\end{rem}

\begin{rem}
Clearly, for Riemannian manifolds, the unit normal vector fields provide
the necessary structure for all $(n-1)$-dimensional submanifolds.
\end{rem}

\section{Boundary Stress and Edge Interactions\label{sec:Edge-Interactions}}

Using the definition of the surface divergence in (\ref{eq:SurfDiv}),
one can now write the boundary integral corresponding to $Y=p_{\st}(X)$
in (\ref{eq:firstIntegrationByParts}) as
\[
\int_{\bdry\body}Y(j^{1}\avf)=\int_{\bdry\body}\dee(\ssp_{\var}(Y)(\avf))-\int_{\bdry\body}\sdiv_{\bdry\body}Y(j^{1}\avf).
\]

If $\bdry\body$ is smooth and $\nor$ is a smooth field, one may
use Stokes's theorem and conclude that
\begin{equation}
\int_{\bdry\body}\dee(\ssp_{\var}(Y)(\avf))=\int_{\bdry(\bdry\body)}\ssp_{\var}(Y)(\avf)=0.
\end{equation}
On the other hand, it will be assumed henceforth that 
\begin{equation}
\bdry\body=\bigcup_{m=1}^{M}\var_{m},
\end{equation}
where each $\var_{m}$ is a smooth $(n-1)$-dimensional submanifold
of $\spc$ so that the intersection
\begin{equation}
E_{lm}=\bdry\var_{l}\cap\bdry\var_{m}
\end{equation}
is either empty or an $(n-2)$-dimensional submanifold with boundary
of $\spc$. We will refer to $E_{lm}$ as the edge between $\var_{l}$
and $\var_{m}$. Naturally, it is assumed that the field $\nor$ is
smooth in each $\var_{m}$ so that the same applies to both $\ssp_{\var_{m}}$
and $\sdiv_{\var_{m}}$. Thus,
\begin{equation}
\begin{split}\int_{\bdry\body}Y(j^{1}u) & =\sum_{m=1}^{M}\int_{\var_{m}}\dee(\ssp_{\var_{m}}(Y)(\avf))-\sum_{m=1}^{M}\int_{\var_{m}}\sdiv_{\var_{m}}Y(j^{1}\avf)\\
 & =\sum_{m=1}^{M}\int_{\bdry\var_{m}}\ssp_{\var_{m}}(Y)(\avf)-\sum_{m=1}^{M}\int_{\var_{m}}\sdiv_{\var_{m}}Y(j^{1}\avf),\\
 & =\sum_{m>l}\int_{E_{ml}}(\ssp_{\var_{m}}(Y)+\ssp_{\var_{l}}(Y))(\avf)-\sum_{m=1}^{M}\int_{\var_{m}}\sdiv_{\var_{m}}Y(j^{1}\avf).
\end{split}
\end{equation}
The integrals over the edges $E_{ml}$ represent edge interactions,
as one would expect.

It follows that Equation (\ref{eq:firstIntegrationByParts}) may be
rewritten for $A=j^{1}\avf$ as 
\[
\begin{split}\int_{\body}X(j^{1}(j^{1}\avf)) & =\sum_{m>l}\int_{E_{ml}}(\ssp_{\var_{m}}(Y)+\ssp_{\var_{l}}(Y))(\avf)-\sum_{m=1}^{M}\int_{\var_{m}}\sdiv_{\var_{m}}Y(j^{1}\avf)\\
 & \qquad-\int_{\body}\diver X(j^{1}\avf).
\end{split}
\]
One observes that since $\diver X$ is a section of $\L{J^{1}\avb,\ext^{n}(T^{*}\spc)}$,
a simple variational stress, we may apply the definition of the generalized
divergence (\ref{eq:DefineDivergence-1}) to it, and so
\begin{equation}
\diver X(j^{1}\avf)=\dee(p_{\st}(\diver X)(\avf))-\diver(\diver X)(\avf).
\end{equation}
Here, similarly to a traction stress $p_{\st}(\diver X)$ is a section
of $\L{\avb,\ext^{n-1}T^{*}\spc}$ and $\diver(\diver X)$ is a section
of $\L{\avb,\ext^{n}(T^{*}\spc)}$, similarly to a body force. Using
(\ref{eq:ReprDivergence-1}) and (\ref{eq:repDivX}), $\diver(\diver X)$,
a section of $L(\avb,\ext^{n}T^{*}\spc)$, is represented locally
by
\begin{equation}
\sum_{\ga,i,j}(X_{1\dots n\ga,ij}^{3ij}-X_{1\dots n\ga,i}^{1i}-X_{1\dots n\ga,i}^{2i}+X_{1\dots n\ga}^{0})g^{\ga}\otimes(\lisup{\dee x}{\wedge}n).
\end{equation}
Equation (\ref{eq:CauchyAndVariational}) implies that $p_{\st}(\diver X)$,
a section of $\L{\avb,\ext^{n-1}T^{*}\spc}$, is represented by 
\begin{equation}
\sum_{i,j,\ga}(-1)^{i-1}(X_{1\dots n\ga,j}^{3ij}-X_{1\dots n\ga}^{1i})g^{\ga}\otimes(\lisupwout{\dee x}{\wedge}ni)
\end{equation}
and its restriction to $\bdry\body$ is represented in chart adapted
to the submanifold in the form (only the term above with $i=n$ does
not vanish in the restriction)
\begin{equation}
\sum_{j,\ga}(-1)^{n-1}(X_{1\dots n\ga,j}^{3nj}-X_{1\dots n\ga}^{1n})g^{\ga}\otimes(\lisupc{dy}{\wedge}{n-1})
\end{equation}
\textemdash a ``body force''-like object on the boundary.

We conclude that, 
\[
\begin{split}\int_{\body}X(j^{1}(j^{1}\avf)) & =\sum_{m>l}\int_{E_{ml}}(\ssp_{\var_{m}}(p_{\st}(X))+\ssp_{\var_{l}}(p_{\st}(X)))(\avf)-\sum_{m=1}^{M}\int_{\var_{m}}\sdiv_{\var_{m}}p_{\st}(X)(j^{1}\avf)\\
 & \qquad\qquad\qquad\qquad\qquad-\int_{\bdry\body}p_{\st}(\diver X)(\avf)+\int_{\body}\diver(\diver X)(\avf).
\end{split}
\]

\newpage{}

\bigskip{}

\noindent \textbf{\textit{Acknowledgments.}} The author is grateful
to F.~dell'Isola for stimulating discussions and comments. This work
was partially supported by the Pearlstone Center for Aeronautical
Engineering Studies and by the H. Greenhill Chair for Theoretical
and Applied Mechanics at Ben-Gurion University.

\end{document}